\documentclass[11pt,a4paper]{emulateapj}

\usepackage{wrapfig}
\usepackage{epsfig}
\usepackage{amsmath}
\usepackage{amssymb}
\usepackage[colorlinks,urlcolor=blue]{hyperref}
\usepackage{breakurl}
\usepackage{xcolor}
\usepackage{graphicx,multirow}
\usepackage{rotating}
\usepackage{placeins}
\usepackage{tikz}
\usetikzlibrary{calc,decorations.pathreplacing}
\usepackage[T1]{fontenc}

\hypersetup{
  colorlinks = true,
  urlcolor = black,
  pdfauthor = {A.\,P.\ Thomson},
  pdfkeywords = {physics},
  pdftitle = {Radio properties of ALMA-identified starbursts -- sizes and radio spectra},
  pdfsubject = {ApJ},
  pdfpagemode = UseNone
}

\newcommand{\aips}{{$\cal AIPS\/$}}

\def\herschel{{\it Herschel}}

\def\s14{$S_{\rm 1.4GHz}$}

\def\jonezero{$J\!=\!1\!-\!0$}

\def\jthreetwo{$J\!=\!3\!-\!2$}

\def\emerlin{$e$-MERLIN}
\newcommand{\co}{$^{12}$CO}
\newcommand{\etal}{et~al.}

\def\gs{\mathrel{\raise0.35ex\hbox{$\scriptstyle >$}\kern-0.6em\lower0.40ex\hbox{{$\scriptstyle \sim$}}}} 
\def\ls{\mathrel{\raise0.35ex\hbox{$\scriptstyle <$}\kern-0.6em\lower0.40ex\hbox{{$\scriptstyle \sim$}}}}

\def\Wm2{\,\hbox{W}\,\hbox{m}^{-2}} 
\def\gsim{\mathrel{\raise0.35ex\hbox{$\scriptstyle >$}\kern-0.6em\lower0.40ex\hbox{{$\scriptstyle \sim$}}}} 
\def\lsim{\mathrel{\raise0.35ex\hbox{$\scriptstyle <$}\kern-0.6em\lower0.40ex\hbox{{$\scriptstyle \sim$}}}} 
\def\ltsima{$\; \buildrel < \over \sim \;$} 
\def\simlt{\lower.5ex\hbox{\ltsima}} 
\def\gtsima{$\; \buildrel > \over \sim \;$} 
\def\simgt{\lower.5ex\hbox{\gtsima}}

\lefthead{Thomson et al.}  \righthead{Radio properties of ALMA-identified starbursts -- sizes and radio spectra}

\begin{document}

\title {Radio spectra and sizes of ALMA-identified Submillimetre Galaxies; evidence of age-related spectral curvature and Cosmic Ray diffusion?}
  
\author{
A.\,P.\ Thomson\altaffilmark{1,2},
Ian Smail\altaffilmark{2},
A.\,M.\ Swinbank\altaffilmark{2},
J.\,M.\ Simpson\altaffilmark{3},
V.\ Arumugam\altaffilmark{4},
S.\ Stach\altaffilmark{2},
E.\,J.\ Murphy\altaffilmark{5},\\
W.\ Rujopakarn\altaffilmark{6,7,8},
O.\,Almaini\altaffilmark{9},
F.\ An\altaffilmark{2},
A.\,W.\ Blain\altaffilmark{11},
C.\,C.\ Chen\altaffilmark{4},
E.\,A.\ Cooke\altaffilmark{2},
U.\ Dudzevi\v{c}i\={u}t\.{e}\altaffilmark{2},
A.\,C.\ Edge\altaffilmark{2},
D.\ Farrah\altaffilmark{12, 13},
B.\ Gullberg\altaffilmark{2},
W.\,Hartley\altaffilmark{14},
E.\ Ibar\altaffilmark{15},
D.\ Maltby\altaffilmark{9},
M.\,J.\ Micha{\l }owski\altaffilmark{10},
C.\ Simpson\altaffilmark{16},\\
P.\ van der Werf\altaffilmark{17},
J.\,L.\ Wardlow\altaffilmark{18}
}
\setcounter{footnote}{0}
\altaffiltext{1}{Jodrell Bank Centre for Astrophysics, The University of Manchester, Oxford Road, Manchester, M13 9PL, UK; email: \href{mailto:alasdair.thomson@manchester.ac.uk}{alasdair.thomson@manchester.ac.uk} }
\altaffiltext{2}{Centre for Extragalactic Astronomy, Department of Physics, Durham University, South Road, Durham DH1 3LE, UK}
\altaffiltext{3}{EACOA fellow: Academia Sinica Institute of Astronomy and Astrophysics, No.\,1, Sec.\,4, Roosevelt Rd., Taipei\,10617, Taiwan}
\altaffiltext{4}{European Southern Observatory, Karl Schwarzschild Strasse 2, Garching, Germany}
\altaffiltext{5}{National Radio Astronomy Observatory, 520 Edgemont Road, Charlottesville, VA\,22903, USA}
\altaffiltext{6}{Department of Physics, Faculty of Science, Chulalongkorn University, 254 Phayathai Road, Pathumwan, Bangkok 10330, Thailand}
\altaffiltext{7}{National Astronomical Research Institute of Thailand (Public Organization), Don Kaeo, Mae Rim, Chiang Mai 50180, Thailand}
\altaffiltext{8}{Kavli Institute for the Physics and Mathematics of the Universe (WPI),The University of Tokyo Institutes for Advanced Study, The University of Tokyo, Kashiwa, Chiba 277-8583, Japan}
\altaffiltext{9}{School of Physics and Astronomy, University of Nottingham, University Park, Nottingham NG7 2RD, UK}
\altaffiltext{10}{University of Leicester, Physics \& Astronomy, University Road, Leicester, LE1 7RH, UK}
\altaffiltext{11}{Astronomical Observatory Institute, Faculty of Physics, Adam Mickiewicz University, ul. Sloneczna 36, 60-286 Pozna\'{n}, Poland}
\altaffiltext{12}{Department of Physics and Astronomy, University of Hawaii, 2505 Correa Road, Honolulu, HI 96822, USA}
\altaffiltext{13}{Institute for Astronomy, 2680 Woodlawn Drive, University of Hawaii, Honolulu, HI 96822, USA}
\altaffiltext{14}{Department of Physics and Astronomy, University College London, 3rd Floor, 132 Hampstead Road, London NW1 2PS, UK}
\altaffiltext{15}{Instituto de F\'isica y Astronom\'ia, Universidad de Valpara\'iso, Avda. Gran Breta\~na 1111, Valpara\'iso, Chile}
\altaffiltext{16}{Gemini Observatory, Northern Operations Center, 670 North A`\={o}h\={o}ku Place, Hilo, HI 96720-2700, USA}
\altaffiltext{17}{Leiden Observatory, Leiden University, P.O. Box 9513, NL-2300 RA Leiden, The Netherlands}
\altaffiltext{18}{Department of Physics, Lancaster University, Lancaster, LA1 4YB, UK}

\begin{abstract}
We analyse the multi-frequency radio spectral properties of $41$ 6\,GHz-detected ALMA-identified, submillimetre galaxies (SMGs), observed at 610\,MHz, 1.4\,GHz, and 6\,GHz with GMRT and the VLA. Combining high-resolution ($\sim0.5''$) 6\,GHz radio and ALMA $870\,\mu$m imaging (tracing rest-frame $\sim20$\,GHz, and $\sim250\,\mu$m dust continuum), we study the far-infrared/radio correlation via the logarithmic flux ratio $q_{\rm IR}$, measuring $\langle q_{\rm IR}\rangle=2.20\pm 0.06$ for our sample. We show that the high-frequency radio sizes of SMGs are $\sim1.9\pm 0.4\times$ ($\sim2$--$3$\,kpc) larger than those of the cool dust emission, and find evidence for a subset of our sources being extended on $\sim 10$\,kpc scales at 1.4\,GHz. By combining radio flux densities measured at three frequencies, we can move beyond simple linear fits to the radio spectra of high-redshift star-forming galaxies, and search for spectral curvature, which has been observed in local starburst galaxies. At least a quarter (10/41) of our sample show evidence of a spectral break, with a median $\langle\alpha^{1.4\,{\rm GHz}}_{610\,{\rm GHz}}\rangle=-0.60\pm 0.06$, but $\langle\alpha^{6\,{\rm GHz}}_{1.4\,{\rm GHz}}\rangle=-1.06\pm 0.04$ obtained via stacking -- a high-frequency flux deficit relative to simple extrapolations from the low-frequency data. We explore this result within this subset of sources in the context of age-related synchrotron losses, showing that a combination of weak magnetic fields ($B\sim35\,\mu$G) and young ages ($t_{\rm SB}\sim40$--$80\,$Myr) for the central starburst can reproduce the observed spectral break. Assuming these represent evolved (but ongoing) starbursts and we are observing these systems roughly half-way through their current episode of star formation, this implies starburst durations of $\lesssim100$\,Myr, in reasonable agreement with estimates derived via gas depletion timescales. 
\end{abstract}


\section{Introduction}

Galaxies selected in the observed-frame $\sim850\,\mu$m window (submillimetre-selected galaxies: hereafter, SMGs) represent a class of extreme star-forming galaxies at cosmological distances. Their rest-frame spectral energy distributions (SEDs) peak in the far-infrared (far-IR), due to the reprocessing of optical/ultraviolet starlight by large column densities of interstellar dust. The far-IR luminosities of SMGs ($L_{\rm IR}\geq 10^{12}$\,L$_\odot$) are similar to those of local Ultra Luminous Infrared Galaxies (ULIRGs), and imply large dust masses and high star-formation rates \citep[$M_{\rm dust}\gtrsim 5\times10^8$\,M$_\odot$, ${\rm SFR}\geq 200$\,M$_\odot$\,yr$^{-1}$, e.g.][]{dacunha15}, while their redshift distribution peaks at $z\sim2$--$3$ \citep[albeit with a significant tail extending to $z>4$, e.g.][]{chapman05, simpson14, brisbin17, danielson17}. At this epoch, SMGs are $\sim1000\times$ more numerous than their low-redshift ULIRG counterparts, and are thought to account for $\sim20$--$40$\% of cosmic star formation \citep[e.g.][]{hughes98, yun12, swinbank14}. The clustering \citep[e.g.][]{hickox12, wilkinson17}, star-formation rates, and large gas reservoirs \citep[e.g.][]{bothwell13} of SMGs  have led to suggestions that they represent a crucial phase in the assembly of local, massive ``red and dead'' elliptical galaxies \citep[e.g.][]{simpson14, toft14, hodge16}.

While much has been learnt about the SMG population since the first submillimetre bolometer observations at the end of the last century \citep[e.g.][]{smail97, hughes98}, a long-standing problem lay in the difficulty of identifying multi-wavelength counterparts to the sources detected in low resolution single-dish submillimetre maps. Exploitation of the relationship between the far-IR and radio emission in star-forming galaxies, allied with the sub-arcsecond resolution of radio interferometers has served as a useful route to identifying SMG counterparts \citep[e.g.][]{ivison98, ivison02}, but recent work with the Atacama Large Millimeter Array (ALMA) -- which allows high-resolution sub-millimetre images to be made -- circumvents the need to probabilistically associate the submillimetre flux seen in single-dish studies with emission in other wavebands \citep[e.g.][]{hodge13,stach19}.

Nevertheless, deep radio observations continue to provide invaluable insight into the nature of SMGs, with lower-frequency ($\nu_{\rm rest}\lesssim 10$\,GHz) observations revealing steep-spectrum ($\alpha\sim-0.8$, where $S_\nu\propto\nu^\alpha$) synchrotron emission (which, in star-forming galaxies is produced predominantly by supernovae, and in galaxies hosting an active galactic nucleus -- AGN -- provides a window on to the central black hole itself), and higher-frequency ($\nu_{\rm rest}\gtrsim 10$\,GHz) observations tracing flatter-spectrum ($\alpha\sim-0.1$) thermal free-free emission, which is believed to arise from the scattering of free-electrons in H{\sc ii} regions around young, massive star clusters \citep[e.g.][]{condon92}. The lack of dust-obscuration in the radio bands ensures that radio observations are as sensitive to dust-obscured star formation (which can also be seen in the far-IR, but generally not in the optical/ultraviolet) as they are to unobscured star formation (which may be seen in the optical/ultraviolet, but not in the far-IR), thus making deep radio imaging an important dust-\textit{unbiased} tracer of star formation \citep[e.g.][]{ivison07}. However, the strong positive $k$-correction in the radio bands makes it increasingly more difficult to detect star formation in galaxies at $z\gtrsim 3$, at which a significant fraction of the SMG population is believed to lie. Moreover, the dual origin of radio emission in galaxies (i.e.\ star-formation and AGN activity) makes the interpretation of radio maps of high-redshift soures dependent on information from other wavebands.

On galaxy-integrated scales, the observed correlation between the far-IR and radio luminosities of star-formation dominated galaxies \citep[the far-IR/radio correlation; ][]{helou85} provides one route toward discriminating dusty starbursts from Compton-thick AGN \citep[e.g.\ ][]{delmoro13}. This correlation spans several orders of magnitude in spatial scale, luminosity, gas surface density and photon/magnetic field density, and owes its existence to the link between both IR and radio emission and the formation and destruction of massive stars. In the simplest ``calorimetry'' models \citep[e.g.\ ][]{lisenfeld96}, the optical/ultraviolet light produced by young, massive stars is absorbed by dust and re-radiated in the far-IR; at the ends of their (short) lives, the supernovae produced by these same stars inject cosmic ray electrons (CREs) into the interstellar medium (ISM), whose eventual energy loss via interaction with the magnetic field of the host galaxy produces synchrotron radio emission. Thus, the far-IR/radio correlation emerges for star-forming systems on timescales longer than the lifetimes of typical OB stars ($\gtrsim 10$\,Myr). Extensive work in samples of higher-redshift galaxies has shown that this correlation broadly holds at out to at least $z\sim 4$ \citep{garrett02,murphy09,bourne11}, with evidence for a modest evolution with redshift \citep[$q_{\rm IR}\propto (1+z)^{-n}$, with $n\lesssim 0.2$:][]{ivison10c, magnelli15, delhaize17, calistrorivera17}.

However at high-redshift, accurate measurements of the radio luminosity densities of galaxies (by convention, measured at rest-frame 1.4\,GHz), depend on a $k$-correction of the observed-frame flux densities to the rest frame, and the magnitude of this $k$-correction is sensitive to the radio spectral index. In a resolution-matched study of 57 Lockman Hole SMGs observed at 610\,MHz and 1.4\,GHz with the Giant Metrewave Radio Telescope (GMRT) and Karl G.\ Jansky Very Large Array (VLA), reaching $1\sigma$ sensitivities of 15 and 6\,$\mu$Jy\,beam$^{-1}$, respectively, \citet{ibar10} measured a median radio spectral index of $\alpha^{1.4\,{\rm GHz}}_{610\,{\rm MHz}}=-0.75\pm 0.06$. Later, in a sample of 52 ALMA-identified SMGs from the LABOCA Extended \textit{Chandra} Deep Field South (ECDFS) Sub-mm Survey (ALMA-LESS -- hereafter the ``ALESS'' sample), \citet{thomson14} measured a median radio spectral index $\alpha^{1.4\,{\rm GHz}}_{610\,{\rm MHz}}=-0.79\pm 0.06$. In both cases, the measured spectral indices were found to be consistent with synchrotron-dominated emission at low radio frequencies. However, some studies have found evidence of spectral-flattening at low frequencies ($\alpha^{1.4\,{\rm GHz}}_{610\,{\rm MHz}}\gtrsim -0.5$) both in local ULIRGs \citep{smith98} and in high-redshift SMGs \citep{hunt05} while others have found evidence of spectral \textit{steepening} at higher frequencies \citep[$\nu\gtrsim 10$\,GHz;][]{galvin18, jimenezandrade19}, in good agreement with models in which the production of secondary electrons and positrons competes with (sometimes rapid) cooling of CREs due to bremsstrahlung, ionization and inverse Compton processes which suppresses high-frequency radio emission \citep{lacki10, basu15}. Together, these works suggest that the rest-frame $\sim 1$--$30$\,GHz radio emission in at least some ULIRGs and SMGs is likely to be more complex than a simple sum of two power laws.

Observations of SMGs with single-dish submillimetre facilities and radio interferometers have shown that, in a galaxy averaged sense, SMGs typically lie on/close to the local far-IR/radio correlation \citep[e.g.\ ][]{ivison10c, thomson14}. Since far-IR and radio emission are both thought -- in the absence of a strong AGN -- to be produced by processes related to star formation, one might anticipate that this relation would hold in SMGs down to the scales probed by our radio and submillimetre observations ($\lesssim 5$\,kpc), as in local dwarf galaxies \citep[e.g.\ ][]{schleicher16}. However, direct comparison of the sizes of the far-IR and radio emission for the same sources has long proved challenging at high-redshift, due to the scarcity of high-resolution, interferometric imaging in the far-IR bands to compare with the deep radio imaging by which counterparts to single-dish submillimetre sources were first identified. Only in the era of submillimetre interferometry ushered-in by the Submillimeter Array (SMA) and ALMA has such a direct comparison between the rest-frame far-IR and radio morphologies of SMGs become possible.

In order to investigate the relationship between the radio and dust continuum emission in SMGs, we have conducted a pilot study of the radio spectral properties of submillimetre galaxies detected in the ALMA survey of the SCUBA-2 Cosmology Legacy Survey UKIDSS/UDS field \citep[hereafter, AS2UDS;][]{stach19}. Using a series of sensitive ($\sigma_{\rm 6\,GHz}\sim 2.7\,\mu$Jy\,beam$^{-1}$), targeted high-resolution ($\sim 0.5''$) C-band ($6$\,GHz) observations obtained with the VLA in A-configuration along with an extremely deep ($\sigma_{\rm 6\,GHz}\sim 0.7\,\mu$Jy\,beam$^{-1}$) two-pointing C-band mosaic made from archival data, we perform a direct comparison of the spatial extents, orientations and morphologies of the radio and dust emission of our SMG targets. By exploiting sensitive VLA L-band (1.4\,GHz) and GMRT 610\,MHz imaging \citep{ibarthesis}, we also gain new constraints on the radio spectral properties of our sources across two intervals in frequency, allowing our analysis to move beyond simple power law characterisations of the radio spectral index, and to search for signs of spectral index curvature.

The remainder of this paper is structured as follows: in \S\,\ref{sect:obs}, we present our sample and observations, including a description of the pre-existing ALMA $870\,\mu$m, VLA 1.4\,GHz and GMRT 610\,MHz observations, as well as a description of the observing, data reduction and imaging strategies used to produce our new VLA 6\,GHz images. In \S\,\ref{sect:results}, we present our results and analysis. We discuss our results in \S\,\ref{sect:discussion}, in which we develop a model whereby both the curved radio spectra and the changes we observe in radio morphology as a function of frequency are explored within the context synchrotron spectral ageing. In \S\,\ref{sect:conclusions} we summarize and offer concluding remarks.

Throughout our manuscript we adopt a $\Lambda$-CDM cosmology with $H_0=71$\,km\,s$^{-1}$\,Mpc, $\Omega_m=0.27$ and $\Omega_\Lambda=0.73$.

\section{Observations and Data Reduction}\label{sect:obs}
\subsection{SCUBA-2/ALMA $870\,\mu$m}
The SMGs in our sample were selected from observations taken as part of the S2CLS survey \citep{geach17} on the James Clark Maxwell Telescope (JCMT). The S2CLS submillimetre map of the UDS field reaches a depth $\sigma_{850}\sim 0.9$\,mJy across $0.96\,{\rm deg}^2$ with a typical beam of $\sim 15''$, and yields $716$ submillimetre sources at $\geq 4\sigma$. In ALMA Cycle 1, \citet{simpson15a} followed up 30 bright ($S_{870\,\mu{\rm m}}\geq8$\,mJy) submillimetre sources, taken from an early version of the S2CLS catalogue at $\sim0.3''$ resolution in Band 7 ($870\,\mu$m). In 30 ALMA pointings, they found 52 SMGs with $S_{\rm 870\,\mu{\rm m}}\geq 1$\,mJy (with a median rms of $\sigma_{870\,\mu{\rm m}}=0.21$\,mJy\,beam$^{-1}$). Details of the pilot ALMA/SCUBA-2 UDS source catalogue, data reduction and imaging can be found in \citet{simpson15a}. The full sample of 716 SCUBA-2 sources -- the AS2UDS sample -- was subsequently observed with ALMA, and is presented in \citet{stach19}.

\subsection{Sample selection and 1.4/6\,GHz radio imaging}
Of the 52 ALMA SMGs studied by \citet{simpson15a}, 29 are detected at $4\sigma\geq 25\,\mu$Jy in deep 1.4\,GHz VLA imaging of the UDS field (V.\ Arumugam \etal, in prep). These 1.4\,GHz observations were carried out under the VLA project AI\,0108, and comprise a mosaic of 14 pointings covering a $\sim 1.3$\,deg$^2$ region, centred on UDS. With $\sim160$\,h total on-source integration time in multiple array configurations (A, B, C, D), the final 1.4\,GHz image reaches a nearly constant rms noise $\sigma\sim 6\,\mu$Jy\,beam$^{-1}$ across the field (as low as $\sigma\sim4\,\mu$Jy\,beam$^{-1}$ near the centre of the mosaic), with a synthesized beam that is well-characterised by a $1.6''$ Gaussian profile. A full description of the observations, data reduction and source catalogue will be presented in V.\ Arumugam \etal\ (in prep).

Using the upgraded VLA between July--September 2015 (Project ID: 15A-249), we conducted a pilot study in A-configuration at C-band towards the 10 SMGs from \citet{simpson15a} with the brightest 1.4\,GHz counterparts. We used the 3-bit receivers with a 2s correlator read time, yielding instantaneous, full-polarization coverage from $4$--$8$\,GHz in $2$\,MHz-wide channels. We hereafter refer to these observations by their central frequency, 6\,GHz. Our observations comprise $70$--$150$\,mins on-source per field. We performed amplitude and bandpass calibration using a single 5\,min scan of 3C\,48 at the beginning of each observing block, and derived phase solutions via a 70\,s scan of the nearby phase reference source, J0215--0222, after each 270\,s scan on the target.

We processed these new 6\,GHz data using the Common Astronomy Software Applications \citep[CASA; ][]{mcmullin07} version 5.1.0 and the included VLA Calibration Pipeline, however post-calibration inspection of the $uv$ data revealed the presence of residual, strong radio frequency interference (RFI), most probably arising from geostationary satellites located in the Clarke Belt, whose declination range intersects the UDS field\footnote{\noindent \url{https://science.nrao.edu/facilities/vla/docs/manuals/obsguide/rfi}}. To mitigate this RFI, we passed the calibrated $uv$ data for each target through the automated {\sc aoflagger} package developed for observations with the Low-Frequency Array \citep[LOFAR; ][]{offringa12}, and then performed a manual search for remaining low-level RFI using the {\sc casa} tool {\sc plotms}. To ease the computational burden of imaging the data without introducing significant smearing effects, we averaged the processed data in time (to an integration time of 3\,s) but not in frequency. We imaged the data from each pointing to the half-power width of the primary beam ($\sim 8'$ per pointing) using {\sc wsclean} \citep{offringa14} with natural weighting and a pixel scale of $0.1''$, which provides 3--5 pixels across the synthesized beam. Finally, we performed primary beam corrections and created image-plane mosaics from overlapping 6\,GHz pointings using the \aips\ task {\sc flatn}.

In addition to these new observations, the VLA has observed a further two deep adjacent pointings at 6\,GHz within the Cosmic Assembly Near-infrared Deep Extragalactic Legacy Survey (CANDELS) region of the UDS field. The first of these pointings was observed under Project ID 12B-175 (PI: Rujopakarn), and comprises approximately 50\,hrs on-source, while the second pointing (15A-048: PI Tadaki), comprises $\sim$20\,hrs on source. The pointing centres of these two images are separated by one half-power beam width. We retrieved the raw $uv$ data for these projects from the VLA archive, and processed, calibrated and imaged them following the same steps as outlined above.

The median rms of our targeted 6\,GHz images is $\sigma_{6\,{\rm GHz}}=4.8\,\mu$Jy\,beam$^{-1}$ ($2.7\,\mu$Jy\,beam$^{-1}$ near the pointing centres), while that of the deep mosaic made from archival data is $\sigma_{6\,{\rm GHz}}=1.6\,\mu$Jy\,beam$^{-1}$ ($0.7\,\mu$Jy\,beam$^{-1}$ near the pointing centres).

While the 15A-249 VLA observations were devised as a follow-up to a sub-set of bright AS2UDS SMGs from the pilot study of \citet{simpson15a}, the subsequent analysis of the full AS2UDS catalogue from \citet{stach19} revealed that 247 ALMA-detected SMGs lie within the combined footprint of our twelve 6\,GHz pointings. Of these 247 SMGs, 41 are detected at $\geq5\sigma_{\rm 6\,GHz}$ via blind source extraction using the {\sc aegean} source finder \citep{hancock12}, where $\sigma$ is the local noise level obtained via boxcar smoothing the 6\,GHz maps. We hereafter refer to these 41 SMGs as our 6\,GHz SMG sample. We show false-colour and radio continuum postage stamps of our 6\,GHz-selected SMG sample in Fig\,\ref{fig:stamps}.

\subsection{GMRT 610\,MHz}\label{sect:gmrt}

To study the low-frequency spectral properties of our SMG targets, we utilise a 610\,MHz image of the UDS field obtained with GMRT. These data were obtained during 2006 February 03--06 and December 05--10, and the details of their reduction -- along with a description of the imaging strategy -- is presented in \citet{ibarthesis} and \citet{dunne09}.

Briefly, this GMRT map was formed from a three-pointing mosaic and comprises a total integration time of 12\,hr per pointing, after setup/calibration overheads.  The observing strategy employed 40\,min scans on the target field, interspersed with 5\,min scans of the bright phase calibrator, 0240--231. Flux and bandpass calibration were performed using the reference sources 3C\,48 and 3C\,147, respectively. Using $128\times1.25$\,kHz channels in each of the two sidebands (centred at 602 and 618\,MHz, respectively) and recording in dual polarization, the final mosaic reaches a typical sensitivity of $\sigma_{\rm 610\,MHz}\sim 60\,\mu$Jy\,beam$^{-1}$ ($\sigma_{\rm 610\,MHz}\sim 40\,\mu$Jy\,beam$^{-1}$ near the centre of the field). The pixel scale of $1.25''$ well-samples the GMRT 610\,MHz synthesized beam ($\theta_{\rm 610\,MHz}\sim 5''$).

\begin{figure*}
\centerline{\psfig{file=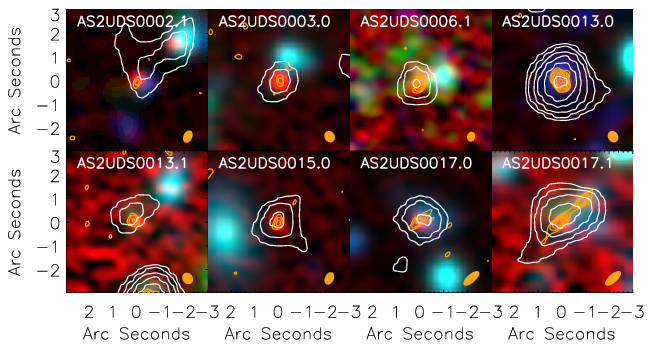,width=\textwidth}}
\vspace*{4mm}
\caption[Short captions]{Postage stamp images of eight representative SMGs with $>5\sigma$\ 6GHz detections from our sample. False colour images are constructed from ALMA $870\,\mu$m (red), and Subaru $i$ (green) and $V$-band (blue) imaging, smoothed with a common $0.35''$ {\sc fwhm} Gaussian kernel to highlight the complex morphology of the stellar continuum emission, and its offsets from the compact regions of dust-enshrouded star-formation seen with ALMA. 6\,GHz (orange) and 1.4\,GHz (white) radio contours are plotted at $-3$, $3$, $3\sqrt{2}\times\sigma$ (and in steps of $\sqrt{2}\times\sigma$ thereafter). The 6\,GHz sizes of our SMGs are on average $\sim (1.8\pm0.4)\times$ larger than the dust continuum sizes measured at $870\,\mu$m (which traces a region $\sim2$--$3$\,kpc in diameter), while at 1.4\,GHz, a number of our sources appear to be marginally-resolved on scales which can be probed by the VLA beam ($1.6''$, corresponding to physical sizes $>10$\,kpc). Thumbnails for the remaining sources are shown in Appendix\,A.1, while the peculiar radio morphology of AS2UDS\,0017.1 is discussed separately in Appendix\,A.4. We show the VLA 6\,GHz synthesized beam as an orange ellipse in the bottom-right corner of each sub-figure.}
\vspace*{4mm}
\label{fig:stamps}
\end{figure*}

On visual inspection of our $870\,\mu$m/radio maps (Fig.\,\ref{fig:stamps} and Appendix\,A.1), it is apparent that a significant number of our 6\,GHz-selected SMGs have companion radio-emitting sources whose separation from the SMG is smaller than $5''$. As a result, the 610\,MHz peak flux densities of these sources will be over-estimated if we do not account for this source confusion. We deblended the GMRT image using the techniques previously outlined in \citet{swinbank14} and \citet{thomson17}. Briefly, we extract a $15''\times15''$ thumbnail around each SMG from the GMRT image, and construct a model 610\,MHz image of the same size which we seed with delta functions at the positions of $870\,\mu$m, 1.4\,GHz and 6\,GHz detected sources (i.e.\ including all likely radio detections within each thumbnail regardless of whether or not they are associated with an SMG). Next, we assign random flux densities to each of the delta functions between $0$--$5\times$ the peak in the GMRT postage stamp and convolve with the GMRT synthesized beam. We create a residual image by subtracting this model from the data, and measure the goodness-of-fit via the $\chi^2$ statistic. We randomly perturb the flux densities assigned to the delta funtions $100,000$ times, or until $\chi^2$ converges on a minimum. For SMGs which lie coincident with a $>3\sigma$ peak in the GMRT image and have no neighbouring radio sources and/or SMGs within the GMRT beam, we measure the 610\,MHz flux density directly from the peak pixel in the (non-deblended) GMRT thumbnail. For SMGs with $>3\sigma$ GMRT emission but nearby radio-detected or SMG companions which could be contributing to the observed flux density, we report flux densities from the corresponding deblended thumbnail. For SMGs which are not coincident with a GMRT source, we report $3\sigma$ upper-limits based on the local noise level.

\begin{table*}
\centering
\caption[Radio/far-IR properties of UDS SMGs -- flux densities and physical sizes]{Radio/far-IR properties of UDS SMGs -- flux densities and physical sizes}
\label{tab:properties}
\begin{tabular}{lccccccccccc}
\hline
\multicolumn{1}{l}{ID$^a$} &
\multicolumn{1}{c}{$S_{6\,{\rm GHz}}^b$} & 
\multicolumn{1}{c}{$S_{1.4\,{\rm GHz}}^c$} &
\multicolumn{1}{c}{$S_{610\,{\rm MHz}}$} &
\multicolumn{1}{c}{$S_{870\,\mu{\rm m}}$} &
\multicolumn{1}{c}{$z_{\rm phot}^d$} & 
\multicolumn{1}{c}{$\theta_{6\,{\rm GHz}}$} & 
\multicolumn{1}{c}{$\theta_{1.4\,{\rm GHz}}$} &
\multicolumn{1}{c}{$\theta_{870\,\mu{\rm m}}$} & \\
\multicolumn{1}{l}{} &
\multicolumn{1}{c}{($\mu$Jy)} &
\multicolumn{1}{c}{($\mu$Jy)} &
\multicolumn{1}{c}{($\mu$Jy)} &
\multicolumn{1}{c}{(mJy)} & 
\multicolumn{1}{c}{} & 
\multicolumn{1}{c}{($''$)} &
\multicolumn{1}{c}{($''$)} &
\multicolumn{1}{c}{($''$)} &\\
\hline
AS2UDS002.1$^\star$&$12\pm2$&$42\pm8$&$<220$&$7.4\pm0.5$&$3.35\pm0.24$&$-$&$-$&$-$\\
AS2UDS003.0$^\star$&$13\pm2$&$50\pm8$&$<200$&$7.9\pm0.4$&$3.93\pm0.91$&$-$&$-$&$0.37\pm0.09$\\
AS2UDS006.1$^\star$&$40\pm8\,\, [6.8]$&$44\pm6$&$<140$&$2.3\pm0.4$&$3.28\pm0.30$&$0.53\pm0.14$&$-$&$0.88\pm0.28$\\
AS2UDS013.0{\color{gray}$^\blacksquare$}&$58\pm4\,\, [25.9]$&$251\pm22\,\, [19.2]$&$350\pm69^{{\color{gray}\blacklozenge}}$&$6.2\pm0.3$&$2.04\pm0.09$&$0.30\pm0.06$&$1.06\pm0.28$&$-$\\
AS2UDS013.1&$12\pm2$&$55\pm10$&$<220$&$1.4\pm0.3$&$2.26\pm0.10$&$-$&$-$&$-$\\
AS2UDS015.0&$15\pm2$&$71\pm10$&$<390$&$5.6\pm0.5$&$6.53\pm0.54$&$-$&$-$&$0.53\pm0.13$\\
AS2UDS017.0&$16\pm2$&$77\pm10$&$<390$&$6.6\pm0.3$&$2.75\pm0.07$&$-$&$-$&$0.46\pm0.07$\\
AS2UDS017.1&$31\pm2$&$224\pm31$&$<390$&$1.5\pm0.4$&$1.26\pm0.15$&$-$&$-$&$-$\\
AS2UDS021.0{\color{gray}$^\blacksquare$}&$136\pm5\,\, [43.5]$&$753\pm30\,\, [40.1]$&$1500\pm70$&$5.5\pm0.3$&$2.26\pm0.06$&$0.35\pm0.03$&$1.35\pm0.12$&$0.63\pm0.06$\\
AS2UDS023.0{\color{gray}$^\blacksquare$}&$22\pm2$&$128\pm16\,\, [12.6]$&$330\pm63$&$6.7\pm0.4$&$2.22\pm0.11$&$-$&$1.23\pm0.39$&$-$\\
AS2UDS039.0&$10\pm2$&$45\pm8$&$<280$&$5.8\pm0.3$&$2.94\pm0.10$&$-$&$-$&$0.47\pm0.08$\\
AS2UDS056.1$^\star${\color{gray}$^\blacksquare$}&$153\pm13$&$401\pm16\,\, [42.3]$&$410\pm77$&$2.0\pm0.6$&$3.17\pm0.07$&$-$&$0.81\pm0.15$&$-$\\
AS2UDS064.0$^\star${\color{gray}$^\blacksquare$}&$126\pm4$&$641\pm20\,\, [52.0]$&$1200\pm90$&$7.4\pm0.8$&$4.15\pm0.40$&$-$&$1.00\pm0.12$&$-$\\
AS2UDS072.0$^\star$&$13\pm3$&$105\pm26\,\, [5.6]$&$<220$&$8.2\pm0.8$&$2.88\pm0.12$&$-$&$2.72\pm0.83$&$-$\\
AS2UDS082.0&$13\pm3$&$36\pm7$&$<270$&$5.2\pm0.5$&$2.58\pm0.07$&$-$&$-$&$0.72\pm0.11$\\
AS2UDS113.0$^\star$&$18\pm4\,\, [6.4]$&$<22$&$<140$&$5.1\pm0.5$&$2.72\pm0.17$&$1.32\pm0.31$&$-$&$0.47\pm0.10$\\
AS2UDS116.0&$18\pm4\,\, [6.2]$&$77\pm15\,\, [7.8]$&$<150$&$6.0\pm0.6$&$2.44\pm0.29$&$0.75\pm0.18$&$1.97\pm0.59$&$0.49\pm0.13$\\
AS2UDS125.0&$30\pm2\,\, [27.8]$&$114\pm13\,\, [14.0]$&$<170$&$4.6\pm0.5$&$1.86\pm0.22$&$0.26\pm0.04$&$1.36\pm0.31$&$-$\\
AS2UDS129.0&$18\pm2$&$<50$&$630\pm140$&$5.2\pm0.7$&$2.75\pm0.29$&$-$&$-$&$0.52\pm0.16$\\
AS2UDS137.0&$26\pm2$&$104\pm8$&$300\pm69$&$5.9\pm0.4$&$2.62\pm0.01$&$-$&$-$&$-$\\
AS2UDS238.0&$35\pm7$&$31\pm8$&$<220$&$4.0\pm0.6$&$2.17\pm0.09$&$-$&$-$&$0.32\pm0.08$\\
AS2UDS259.0&$33\pm6\,\, [8.4]$&$120\pm13\,\, [14.5]$&$<140$&$4.7\pm0.3$&$1.86\pm0.04$&$0.47\pm0.13$&$1.27\pm0.29$&$0.33\pm0.09$\\
AS2UDS265.0&$10\pm2$&$43\pm7$&$<240$&$3.7\pm0.6$&$2.30\pm0.07$&$-$&$-$&$-$\\
AS2UDS266.0&$7\pm1$&$37\pm6$&$<130$&$4.2\pm0.7$&$2.75\pm0.25$&$-$&$-$&$-$\\
AS2UDS272.0{\color{gray}$^\blacksquare$}&$90\pm2\,\, [74.0]$&$260\pm12\,\, [37.4]$&$220\pm54$&$5.1\pm0.5$&$1.78\pm0.21$&$0.40\pm0.01$&$0.70\pm0.20$&$-$\\
AS2UDS283.0{\color{gray}$^\blacksquare$}&$24\pm5$&$116\pm22\,\, [7.9]$&$280\pm67$&$3.9\pm0.7$&$1.88\pm0.12$&$-$&$2.02\pm0.58$&$0.85\pm0.21$\\
AS2UDS297.0{\color{gray}$^\blacksquare$}&$27\pm3\,\, [15.3]$&$104\pm15\,\, [11.0]$&$200\pm49$&$4.4\pm0.6$&$1.68\pm0.20$&$0.46\pm0.07$&$1.91\pm0.42$&$-$\\
AS2UDS305.0&$13\pm2\,\, [7.4]$&$30\pm7$&$<160$&$4.7\pm0.3$&$2.88\pm0.32$&$0.59\pm0.14$&$-$&$0.31\pm0.08$\\
AS2UDS311.0&$21\pm3\,\, [8.4]$&$57\pm14\,\, [6.2]$&$<140$&$5.8\pm0.8$&$2.14\pm0.10$&$0.59\pm0.13$&$1.94\pm0.64$&$0.54\pm0.17$\\
AS2UDS407.0{\color{gray}$^\blacktriangle$}&$15\pm2$&$54\pm7$&$<260$&$3.3\pm0.7$&$2.16\pm0.24$&$-$&$-$&$-$\\
AS2UDS412.0&$17\pm3\,\, [6.0]$&$31\pm6$&$<140$&$4.1\pm0.3$&$2.60\pm0.19$&$0.74\pm0.19$&$-$&$0.47\pm0.09$\\
AS2UDS428.0{\color{gray}$^\blacksquare$}&$98\pm6\,\, [28.2]$&$404\pm18\,\, [37.9]$&$560\pm69$&$4.7\pm0.8$&$1.67\pm0.04$&$0.26\pm0.06$&$0.90\pm0.14$&$0.55\pm0.16$\\
AS2UDS460.1{\color{gray}$^\blacksquare$}&$19\pm2$&$131\pm18\,\, [11.9]$&$370\pm84$&$3.1\pm0.7$&$2.74\pm0.16$&$-$&$1.28\pm0.36$&$-$\\
AS2UDS483.0&$17\pm2$&$128\pm23\,\, [8.7]$&$<390$&$3.1\pm0.3$&$1.86\pm0.33$&$-$&$1.75\pm0.39$&$0.47\pm0.09$\\
AS2UDS497.0&$31\pm2$&$142\pm7$&$340\pm75$&$2.4\pm0.2$&$0.74\pm0.01$&$-$&$-$&$0.36\pm0.11$\\
AS2UDS550.0$^\star$&$19\pm3$&$50\pm6$&$<210$&$4.9\pm0.5$&$3.05\pm0.17$&$-$&$-$&$0.62\pm0.12$\\
AS2UDS590.0&$9\pm2$&$<26$&$<210$&$3.3\pm0.3$&$2.42\pm0.11$&$-$&$-$&$0.42\pm0.11$\\
AS2UDS608.0$^\star${\color{gray}$^\blacktriangle$}&$50\pm9\,\, [8.1]$&$144\pm9$&$<210$&$3.5\pm0.4$&$2.47\pm0.13$&$0.55\pm0.16$&$-$&$-$\\
AS2UDS648.0&$14\pm2$&$38\pm6$&$<190$&$1.8\pm0.5$&$2.48\pm0.05$&$-$&$-$&$-$\\
AS2UDS665.0&$11\pm2$&$32\pm7$&$<200$&$2.3\pm0.3$&$2.10\pm0.26$&$-$&$-$&$-$\\
AS2UDS707.0$^\star${\color{gray}$^\blacktriangle$}&$29\pm6\,\, [6.2]$&$40\pm7$&$<290$&$2.2\pm0.3$&$2.53\pm0.15$&$0.58\pm0.14$&$-$&$-$\\
\hline
Stack (all)&$22\pm1$&$92\pm5$&$181\pm17$&$4.7\pm0.3$&--&$0.54\pm0.04$&$1.36\pm0.16$&$0.28\pm0.06$\\
Stack (Bright)&$21\pm5$&$152\pm6$&$291\pm18$&$5.1\pm0.4$&--&$0.63\pm0.04$&$1.10\pm0.13$&$0.22\pm0.05$\\
Stack (Faint)&$20\pm5$&$49\pm5$&$93\pm12$&$4.2\pm0.4$&--&$0.49\pm0.08$&$1.26\pm0.29$&$0.33\pm0.07$\\
Stack (Convex)&$56\pm2$&$262\pm7$&$430\pm24$&$5.1\pm0.1$&--&$0.51\pm0.03$&$1.03\pm0.14$&$0.20\pm0.04$\\
\hline
\end{tabular}
\\ {\small Notes: $^a$IDs follow those of \citet{stach19}, and differ from those of \citet{simpson15a} which were based on a preliminary version of the AS2UDS sample; $^{b,c}$For unresolved sources we report peak flux densities and uncertainties. For spatially-resolved sources we report the fitted (integrated) flux densities and uncertainties, and report the peak S/N in square brackets for reference; $^d$Photometric redshifts are obtained via multi-band SED fits to the UKIDSS Ultra-Deep Survey DR11 catalogue (O.\ Almaini \etal, in prep; U.\ Dudzevi\v{c}i\={u}t\.{e} \etal, in prep);  $^\star$Candidate AGN host based on \textit{Spitzer} IRAC colours; $^{\color{gray}\blacktriangle}$Candidate AGN host based on X-ray detection; $^{\color{gray}\blacklozenge}$GMRT flux density measured from deblended thumbnail; {\color{gray}$^\blacksquare$}Source is detected in all three radio bands and has a convex spectrum, i.e. $\alpha^{\rm 1.4\,GHz}_{\rm 610\,MHz}>\alpha^{\rm 6\,GHz}_{\rm 1.4\,GHz}$. }
\end{table*}

\subsection{Size measurements}\label{sect:size_methods}

We measure deconvolved angular sizes for our SMGs by fitting two-dimensional Gaussian models in the 1.4\,GHz and 6\,GHz radio maps at the positions of the SMGs using the {\sc casa} task {\sc imfit}. We report these sizes in Table\,\ref{tab:properties}. At the resolution of our GMRT map, we do not expect any of our SMGs to be resolved ($\theta_{\rm 610\,MHz}\sim 5''$ corresponds to a linear scale of $\sim 40$\,kpc at $z\sim 2$), and so we do not perform forced Gaussian fitting to the 610\,MHz map.

\section{Results \& Analysis}\label{sect:results}

\subsection{Radio flux densities, and spectral indices}\label{sect:results_alpha}

We create maps of the local rms and background from our three-band radio data using a custom {\sc idl} script which boxcar-smooths the maps, and perform blind source extraction on the 1.4\,GHz and 6\,GHz images using the {\sc aegean} source finder \citep{hancock12}. We employ a local $5\sigma$ threshold to define peaks in the image, and then perform 2D Gaussian fits to these peaks using the {\sc casa} task {\sc imfit}, yielding both total/peak flux densities and fitted/deconvolved source sizes. Next, we isolate the radio counterparts to SMGs by cross-matching the resulting 1.4\,GHz and 6\,GHz catalogues to the AS2UDS $870\,\mu$m catalogue with a $1''$ search radius. For SMGs which lack a radio counterpart in one or more bands, we measure a $3\sigma$ upper-limit to the radio flux density (assuming the source is unresolved) from the corresponding local rms map. As discussed in \S\,\ref{sect:gmrt}, we measure GMRT flux densities from the peak pixel value at the position of the SMG using either the raw or deblended image, depending on the number of probable confusing sources nearby.

The three-band radio flux densities (and upper-limits) of our 6\,GHz-selected SMGs are reported in Table\,\ref{tab:properties}, with the measured spectral indices (or spectral index limits, in the case of sources which are undetected in one of the two radio bands) shown in Table\,\ref{tab:diffusion}. We measure the spectral indices in two frequency ranges, from 610\,MHz to 1.4\,GHz ($\alpha^{1.4\,{\rm GHz}}_{610\,{\rm MHz}}$) and from 1.4\,GHz to 6\,GHz ($\alpha^{6\,{\rm GHz}}_{1.4\,{\rm GHz}}$). The median 1.4\,GHz flux density of our sample is $\langle S_{1.4,{\rm GHz}}\rangle =92\pm5\,\mu$Jy.

Twelve of our SMGs are detected in all three radio bands, with a median spectral index $\langle \alpha^{1.4\,{\rm GHz}}_{610\,{\rm MHz}}\rangle = -0.80\pm0.14$, which is consistent with the typical spectral indices seen at these frequencies in previous SMG studies \citep[e.g.\ ][]{ibar10, thomson14}, and with measurements of the (synchrotron-dominated) low-frequency radio spectral indices in local starbursts \citep[e.g.\ M82: ][]{condon92} and low-SFR high-redshift sources \citep{murphy17}.

\begin{figure*}
\centerline{\psfig{file=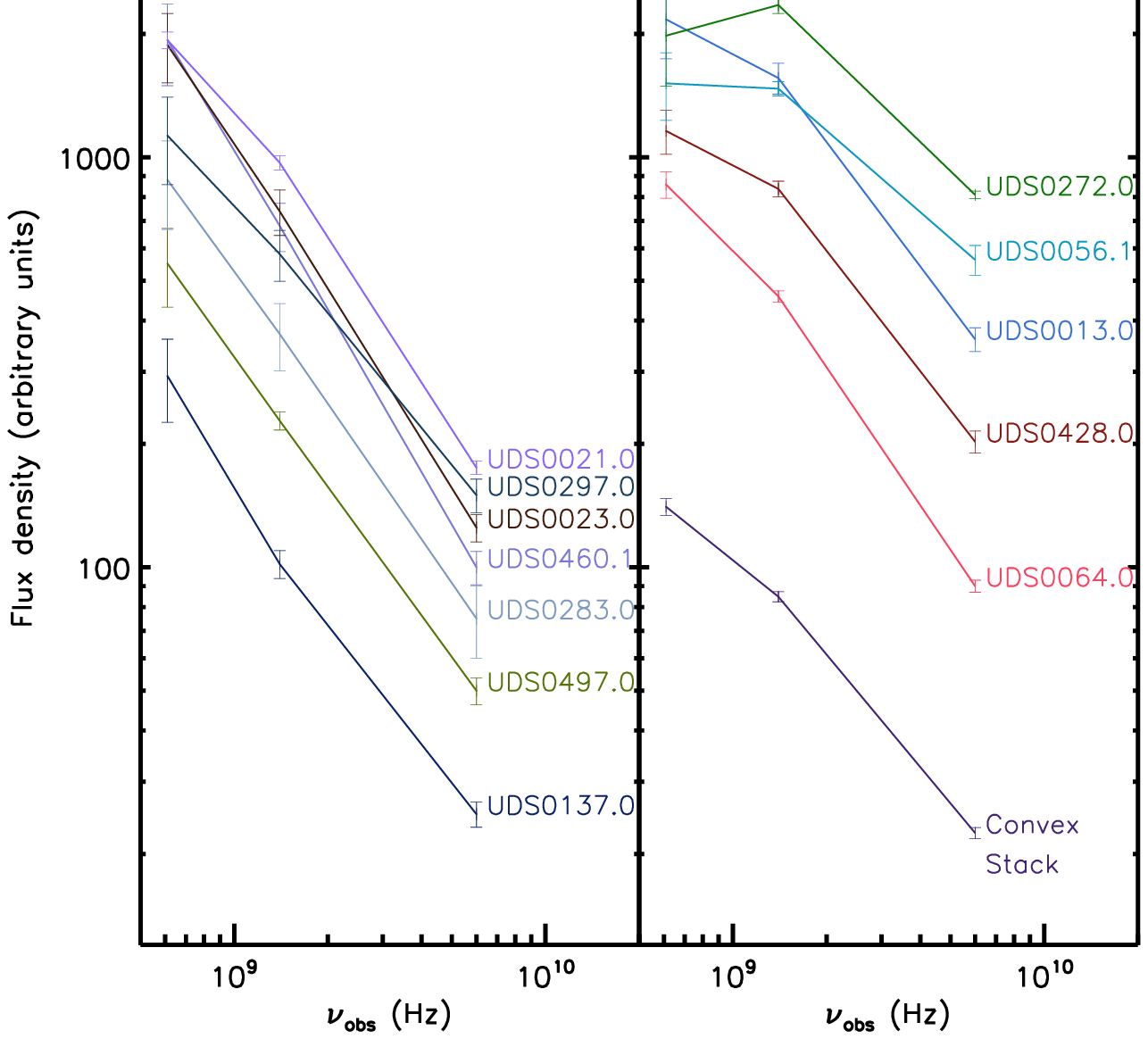,height=14cm,angle=90}}
\vspace*{4mm}
\caption[]{Observed-frame radio SEDs for the 12 SMG targets detected at 610\,MHz, 1.4\,GHz and 6\,GHz, ranked by the strength of the observed spectral break, $\lvert \alpha^{6{\rm GHz}}_{1.4\,{\rm GHz}}-\alpha^{1.4{\rm GHz}}_{610\,{\rm MHz}}\rvert$, with the seven SMGs showing the weakest break shown in the left panel and the six SMGs showing the strongest break shown in the right panel. The SEDs have been arbitrarily re-normalised in flux to allow them to be plotted on the same panels, facilitating a comparison of their spectral shapes. Prior to conducting the 6\,GHz observations, our expectation was that the radio SEDs would flatten at higher frequency due to the increasing contribution of thermal free-free emission at $\nu_{\rm rest}\gtrsim 10$\,GHz \citep[e.g.\ ][]{condon92, murphy17}, however we fail to observe significant spectral flattening in 10/12 SMGs with sufficient radio luminosities to be detected in all three radio bands. To illustrate that this apparent spectral-steepening (or lack of expected spectral-flattening) is not simply driven by low signal-to-noise maps, we also include a stacked SED of these 10 SMGs in the right panel (the ``convex'' SMG sample, which has $\alpha^{\rm 6\,GHz}_{\rm 1.4\,GHz}=-1.06\pm0.04$ and $\alpha^{\rm 1.4\,GHz}_{\rm 610\,MHz}=-0.60\pm 0.06$).}
\vspace*{4mm}
\label{fig:bright13}
\end{figure*}

Turning to higher frequencies, the 41 SMGs detected at 6\,GHz have a median flux density $S_{6\,{\rm GHz}}=22\pm1\,\mu$Jy. Prior to the analysing the maps, we anticipated that the 6\,GHz flux densities would be $\sim 50$\% higher than this, owing to the combination of synchrotron emission (extrapolated from their previously-measured 1.4\,GHz flux densities assuming a typical spectral index $\alpha_{\rm sync}=-0.8$) plus thermal free-free emission, which we expected to contribute an additional $\sim 10$--$20\,\mu$Jy, given the high star-formation rates estimated from the far-IR SED fits (${\rm SFR}_{\rm IR}\sim 500$\,M$_\odot$\,yr$^{-1}$), leading to a predicted $\alpha^{6\,{\rm GHz}}_{1.4\,{\rm GHz}}\gtrsim -0.5$. In fact, the flux densities of our 6\,GHz sample result in a median high-frequency spectral index $\langle \alpha^{6\,{\rm GHz}}_{1.4\,{\rm GHz}}\rangle=-0.98\pm0.07$, which is slightly steeper than the low-frequency spectral index for the whole sample, and suggests there is no evidence of the expected flattening of the spectrum at higher-frequency due to thermal free-free emission. Similar spectral behaviour has also recently been reported in $310$\,MHz--$3$\,GHz observations undertaken by the VLA-COSMOS 3\,GHz Large Project \citep{tisanic19}, who attribute the effect to lower-than-expected thermal free-free emission \citep[see also][]{barcosmunoz15}.

Given this somewhat unexpected finding, we performed a series of consistency checks to test the accuracy of our flux density measurements, using both the processed VLA $uv$ data and also simulated $uv$ datasets representing ``observations'' of model galaxies of known size/flux density under the same conditions as for the real observations. Details of these tests are given in Appendix\,A.2. In summary, we find no evidence that the lower-than-expected 6\,GHz flux densities are the result of either instrumental effects, or systematic problems with the calibration of the flux density scale.

\subsection{Stacking analysis}\label{sect:stacking}

To check that the observed spectral behaviour is not a spurious result driven by low signal-to-noise detections, we perform a stacking analysis. We stack our 1.4\,GHz and 610\,MHz data in the image plane by extracting $20''$ thumbnails around each SMG. We resample these thumbnails 100 times, measuring the median flux density in each pixel, and then create final, stacked thumbnails by computing the ``median of the medians'' from these 100 stacked sub-samples. At 610\,MHz, we extract thumbnails from the published map \citep{ibarthesis} for SMGs with no 1.4\,GHz or SMG companions within $5''$ (corresponding to the GMRT synthesized beam). For SMGs with nearby companions that may be contributing to the 610\,MHz flux density at the position of the SMG, we use thumbnails extracted from the deblended model image (see \S\,\ref{sect:gmrt}) with the nearby sources removed.

In addition to creating median stacks, we also create error images for each of the stacks by computing in each pixel the standard deviation of the bootstrap-resampled thumbnails used in the stacking procedure. We measure uncertainties in our stacked flux densities from these maps by measuring the peak pixel value within $1''$ of the centroid of the error image.

At 6\,GHz and 870\,$\mu$m, our sources are not cropped from a single wide-field image with a stable point spread function (PSF) but were observed in multiple pointings, over a range of elevations and with different beam shapes. As a result, simple image-plane stacking of the kind performed in the 610\,MHz and 1.4\,GHz maps would be inappropriate, as the flux density units of our maps are Jy\,beam$^{-1}$, and the beam varies from pointing-to-pointing. We therefore employ the {\sc stacker} library developed for use in {\sc casa} \citep{lindroos15} to generate median 6\,GHz and 870\,$\mu$m stacks in the $uv$ plane, from which we then create stacked images using {\sc casa} {\sc tclean}; by stacking the data in the $uv$ plane and then performing a single imaging run (with a single, well-defined PSF) on the gridded, stacked $uv$ data we are able to circumvent issues which would otherwise arise from the inhomogeneous PSFs of our individual 6\,GHz and 870\,$\mu$m maps.

\begin{figure*}
\centerline{\psfig{file=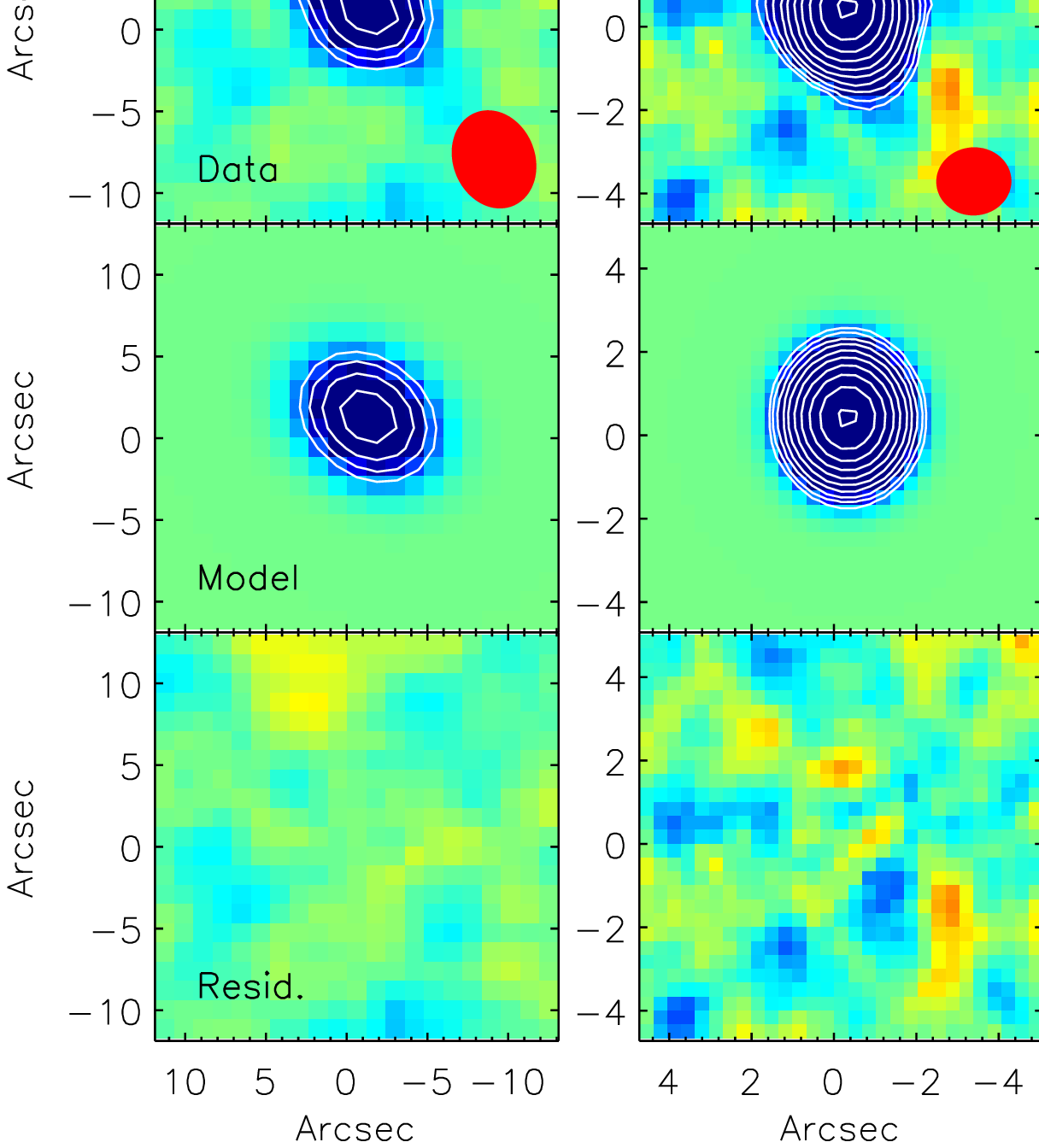,width=19cm}}
\vspace*{-4mm}
\caption[Short captions]{\textit{Top:} False-colour stacked thumbnail images at 610\,MHz, 1.4\,GHz, 6\,GHz and 870\,$\mu$m for the sample of 10 SMGs exhibiting a clear break in their radio spectra at $\sim$GHz frequencies (i.e.\ the ``convex'' sample), with white contours over-plotted at $-3$,$3$,$\sqrt(2)3\times\sigma$ (and in steps of $\sqrt(2)\times\sigma$ thereafter), where $\sigma$ is the local rms within each thumbnail image ($\sigma_{\rm 610\,MHz}=37\,\mu$Jy\,beam$^{-1}$; $\sigma_{\rm 1.4\,GHz}=3.5\,\mu$Jy\,beam$^{-1}$; $\sigma_{\rm 6\,GHz}=0.8\,\mu$Jy\,beam$^{-1}$; $\sigma_{{\rm 870}\,\mu{\rm m}}=57\,\mu$Jy\,beam$^{-1}$). Note the difference in the angular sizes of these thumbnails; we show the PSF corresponding to each image as a red filled circle in the bottom-right of the sub-figures in the top row. The colour scale of the thumbnail images runs between $\pm5\sigma$. \textit{Middle:} Single-component 2D Gaussian model fits to the thumbnail images obtained using the {\sc casa} task {\sc imfit}, shown with the same colour stretch and contour spacing as for the data. Note that the model images are {\bf not} deconvolved from the PSF, however the angular sizes reported in Table\,\ref{tab:properties} {\bf are}. Unsurprisingly, at the angular resolution of our 610\,MHz map ($\sim 5''$) the best-fit model is an unresolved point source. At 1.4\,GHz, however, the angular resolution ($\sim 1.6''$) and high signal-to-noise of our stacked thumbnail (S/N$\sim 35$) allows a resolved source model to be fit with a deconvolved major axis $\theta_{\rm 6GHz}\sim 1.03\pm 0.14''$. Likewise at 6\,GHz and 870\,$\mu$m we measure deconvolved source sizes of $\theta_{\rm 6\,GHz}=0.51\pm 0.03''$ and $\theta_{{\rm 870}\,\mu{\rm m}}=0.20\pm 0.04''$, from images with peak S/N$\sim 21$ and $\sim 25$, respectively. \textit{Bottom:} Residual images (i.e.\ data$-$model) for the four stacked thumbnail images, again shown with the same colour stretch and contour spacing as for the original thumbnails. The 610\,MHz and 1.4\,GHz residual thumbnail images are noise-like, as are the 6\,GHz and 870\,$\mu$m residual thumbnail images (save for a single beam-sized $3\sigma$ peak in the former image and four sub-beam sized $3\sigma$ peaks in the latter image), highlighing that the single-component Gaussian fits (and the sizes measured from them) well-characterise the 2D flux distributions of our convex stacked sample.}
\vspace*{4mm}
\label{fig:stack_stamps}
\end{figure*}

In order to search for evolution in the spectral properties of our SMGs as a function of radio flux density, we generate stacks for the entire sample of 41 6\,GHz-detected SMGs, as well as for samples comprised of SMGs above and below the median 1.4\,GHz flux density, which we label the ``all'', ``bright'' and ``faint'' stacks, respectively. In addition, to further investigate the unexpected spectral index curvature seen in the 10/12 SMGs with detections in three radio bands, we create an additional stack comprised of those SMGs with $\alpha^{\rm 1.4\,GHz}_{\rm 610\,MHz}>\alpha^{\rm 6\,GHz}_{\rm 1.4\,GHz}$. We label this the ``convex'' subsample. We show individual SEDs for the 12 SMGs detected in all three radio bands in Fig.\,\ref{fig:bright13}, along with the convex stacked SED, and show stacked thumbnail images from the convex sample in Fig.\,\ref{fig:stack_stamps}.

The bright and faint stacked sub-samples have high-frequency spectral indices of $\alpha^{\rm 6\,GHz}_{\rm 1.4\,GHz}=-1.35\pm0.24$ and $\alpha^{\rm 6\,GHz}_{\rm 1.4\,GHz}=-0.81\pm0.11$ and low frequency spectral indices of $\alpha^{\rm 1.4\,GHz}_{\rm 610\,MHz}=-0.79\pm0.07$ and $\alpha^{\rm 1.4\,GHz}_{\rm 610\,MHz}=-0.77\pm0.16$, respectively, suggesting that brighter SMGs may have intrinsically steeper spectra between rest-frame $\sim 3$--$20$\,GHz than they do at lower frequenies. For the convex subset (representing around $\sim 25$\% of our 6\,GHz SMG sample) we measure $\alpha^{\rm 6\,GHz}_{\rm 1.4\,GHz}=-1.06\pm 0.04$ and $\alpha^{\rm 1.4\,GHz}_{\rm 610\,MHz}=-0.60\pm 0.06$, a $\gtrsim 4\sigma$ difference in the low and high-frequency spectral indices in this subsample\footnote{We note that by definition all 41 6\,GHz-selected SMGs are detected at 6\,GHz, of which the majority (38/41) are also detected at 1.4\,GHz. The 610\,MHz detection rate is 13/41, which suggests that any potential issues due to flux-boosting from low S/N detections are more likely to affect $S_{\rm 610\,MHz}$ than $S_{\rm 6\,GHz}$. If $S_{\rm 610\,MHz}$ is systematically over-estimated due to flux-boosting effects, then this would artificially \textit{reduce} the strength of the spectral break rather than cause it.}. If the radio emission at all three frequencies shares a common origin (i.e.\ synchrotron and free-free emission from current star-formation) then the implication of this spectral steepening is that either the synchrotron or free-free components (or conceivably, both) are suppressed at higher frequency, relative to simple extrapolations from lower-frequency emission. Alternatively, the high- and low-frequency radio emission in these SMGs may arise from decoupled processes, in which case the curvature seen in the source-integrated radio SEDs may arise from the mixing of emission from proceses that dominate in different frequency ranges and potentially on different physical scales. We will return to this idea in \S\,\ref{sect:cr-propagation}.

\begin{figure}
   \centerline{\psfig{file=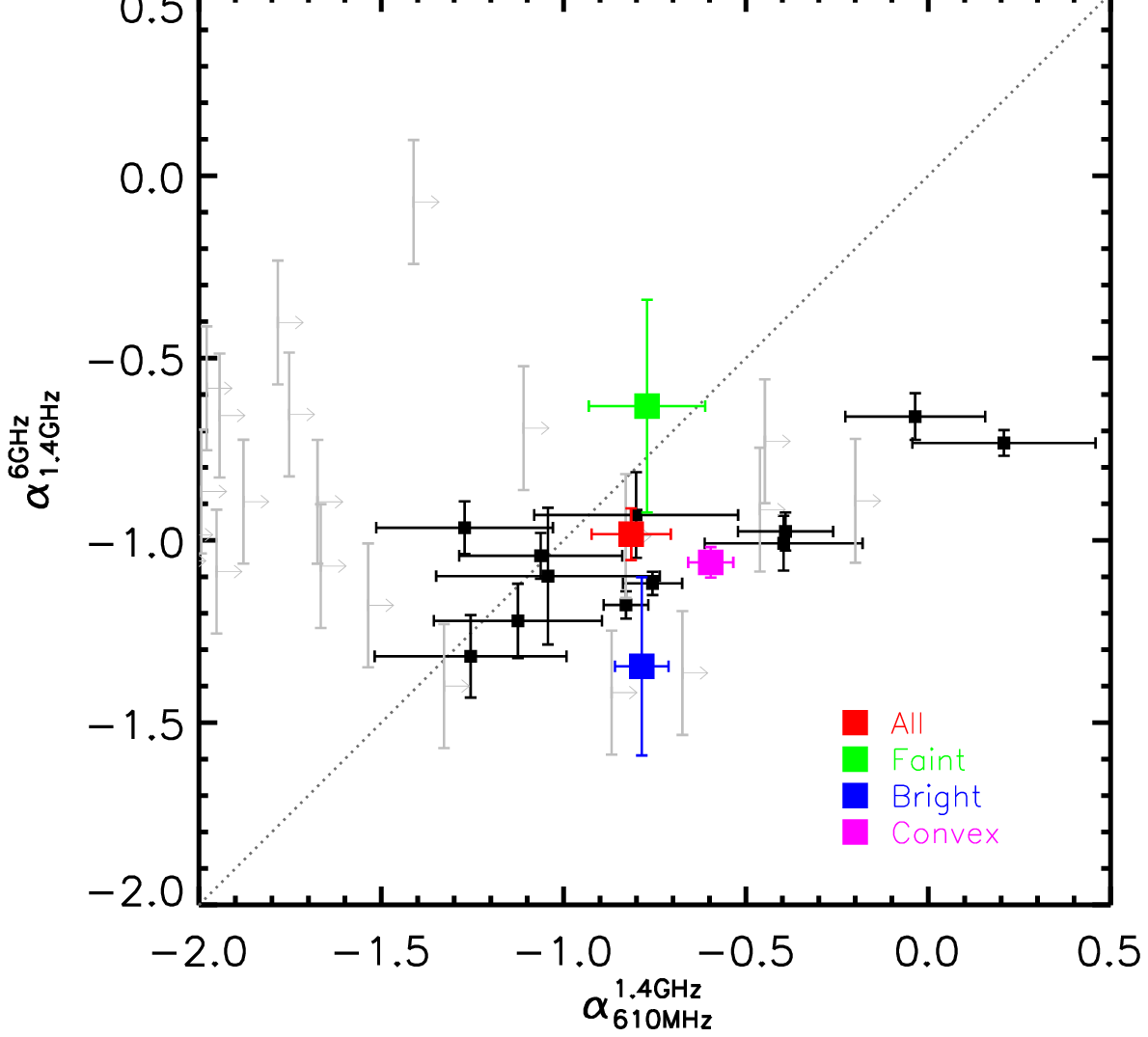,height=\columnwidth,angle=90}}
  \caption[Short captions]{A comparison between the high- and low-frequency radio spectral indices of our 6\,GHz-selected SMG sample, including the 13 SMGs with detections in all three radio bands, 26 with detections at 1.4\,GHz and 6\,GHz (i.e.\ with measured high-frequency spectral indices but only $3\sigma$ limits on their low-frequency spectral indices) and each of the stacked sub-samples outlined in \S\,\ref{sect:stacking}. In local starburst galaxies and $\sim\,\mu$Jy radio galaxies at high-redshift, steep-spectrum synchrotron emission dominates at low frequency and gives way at higher-frequency to the flatter-spectrum free-free emission \citep[e.g.\ ][]{condon92, murphy17}. In this scenario, we would expect star-forming galaxies to lie above the 1:1 line. Instead, for those SMGs that are bright enough to have their spectral indices measured across two frequency intervals, we see a tendency for their SEDs to \textit{steepen} at higher frequency.}
\vspace*{4mm}
\label{fig:alpha_v_alpha}
\end{figure}

The measured low- and high-frequency spectral indices of our sample (including stacked subsamples) are shown in Fig\,\ref{fig:alpha_v_alpha}.

\subsection{The far-infrared/radio correlation}\label{sect:qir}

To measure the rest-frame radio luminosities ($L_{\rm 1.4\,{\rm GHz}}$) of our sample, we must first $k$-correct the observed-frame 1.4\,GHz flux densities:

\begin{center}
  \begin{equation}
    L_{\rm 1.4\,{\rm GHz, rest}}\equiv L_{\rm 1.4\,{\rm GHz}}=4\pi D_{L}^{2} S_{1.4\,{\rm GHz, obs}} (1+z)^{-1-\alpha}
  \end{equation}
\end{center}

\noindent where $D_{L}$ is the luminosity distance to the source, and the subscripts ``rest'' and ``obs'' denote rest-frame and observed-frame quantities, respectively.

Our three-band radio photometry provides independent spectral indices on either side of $\nu_{\rm obs}=1.4$\,GHz. Emission from \textit{rest-frame} 1.4\,GHz in a $z\sim 2.3$ galaxy is shifted to lower frequencies ($\sim 400$\,MHz), while emission at \textit{observed} 1.4\,GHz was originally emitted at higher frequency in the rest-frame. Therefore to obtain rest-frame 1.4\,GHz flux densities from observed-frame $S_{1.4\,{\rm GHz}}$ in the presence of spectral curvature, the appropriate spectral index to use is $\alpha^{\rm 1.4\,GHz}_{\rm 610\,MHz}$. Because the majority of our 6\,GHz SMG sample lack a $>3\sigma$ detection in the GMRT 610\,MHz image, we can only set lower-limits on  $\alpha^{\rm 1.4\,GHz}_{\rm 610\,MHz}$ for these sources. Where these lower-limits are consistent with the sample median ($\alpha^{\rm 1.4\,GHz}_{\rm 610\,MHz}=-0.84\pm0.10$), we $k$-correct using this spectral index, and where the $3\sigma$ GMRT flux limits necessitate a flatter spectral index than the sample median, we $k$-correct using the corresponding $3\sigma$ spectral index limit (Table\,\ref{tab:diffusion}.)

In U.\ Dudzevi\v{c}i\={u}t\.{e} \etal (in prep), we measure the photometric redshifts of our SMGs via SED fits to the multi-band photometry in the UDS field ($UBVRIzYJHK$, InfraRed Array Camera (IRAC) $3.6$, $4.5$, $5.8$, $8.0\,\mu$m, MIPS $24\,\mu$m, \herschel\ PACS $100$, $160\,\mu$m, deblended SPIRE $250\,\mu$m, $350\,\mu$m, $500\,\mu$m, ALMA $870\,\mu$m and VLA 1.4\,GHz) obtained using the {\sc magphys} code \citep{dacunha08}. {\sc magphys} employs the stellar population synthesis models of \citet{bruzual03} with a \citet{chabrier03} stellar initial mass function (IMF) combined with a two-component dust attenuation model \citep{charlot00}, balancing the energetics between the mid- and far-IR dust components to disentangle the integrated dust-attenuated stellar emission of the galaxy and the dust-reprocessed stellar emission. From these SED fits we also obtain rest-frame $8$-$1000\,\mu$m luminosities, $L_{\rm IR}$. Full details of the {\sc magphys} SED fitting and the resulting multiwavelength properties will be described in U.\ Dudzevi\v{c}i\={u}t\.{e} \etal (in prep). We now use these measurements of $L_{\rm IR}$ in conjunction with the rest-frame radio luminosities to study the far-IR/radio correlation, via the parameter:

\begin{center}
  \begin{equation}
    q_{\rm IR} = \log\biggl[\frac{L_{\rm IR}}{3.75\times10^{12}\,{\rm W}} \times \frac{{\rm W\,Hz}^{-1}}{L_{1.4\,{\rm GHz}}}\biggr]
  \end{equation}
\end{center}
    
\noindent as in \citep{ivison10c}.

We measure a median $\langle q_{\rm IR}\rangle=2.20\pm0.06$ for the 6\,GHz and 1.4\,GHz-detected SMGs (Table\,\ref{tab:diffusion}), a little lower than that measured in a sample of 52 radio-detected ALMA SMGs from the ECDFS field \citep[$q_{\rm IR}=2.56\pm0.05$; ][]{thomson14}, a result which is likely driven by our 6\,GHz selection criterion.

Four of our SMGs have $q_{\rm IR}< 1.7$, which marks the classical cut-off between star-formation-dominated and ``radio-excess'' (i.e.\ AGN-dominated) sources \citep[e.g.\ ][]{delmoro13} -- however, none of these SMGs has a bright X-ray counterpart, either in the $1.3$\,deg$^2$ \textit{XMM} SXDS catalogue \citep{ueda08}, or in the deeper \textit{Chandra} coverage \citep{kocevski18}, and thus if these radio-excess sources host active nuclei, they are likely to be Compton thick. Ten of our SMGs have mid-IR colours consistent with a dusty torus \citep[e.g.\ ][]{donley12}, of which two also fit the radio excess ($q_{\rm IR}\lesssim 1.7$) criterion. 15/41 SMGs ($\sim 37\%$ of our sample) satisfy at least one of the radio excess, X-ray detection or IRAC colour-colour criteria, highlighing them as possible AGN, however we note that dust-reddened stellar SEDs of high-redshift starburst galaxies can be mistaken for AGN in simple colour classification schemes (e.g.\ Radcliffe et al, submitted). No sources meet all three AGN criteria, however three (AS2UDS\, 064.0, 608.0 and 707.0) meet two AGN criteria. In addition to $q_{\rm IR}$ values, we also highlight in Table\,\ref{tab:diffusion} which sources meet the IRAC colour-colour criterion for AGN and which are X-ray detected.

\subsection{The sizes \& morphologies of SMGs}\label{sec:sizes}

We now compare our measured radio sizes with the $870\,\mu$m dust sizes of our sample from \citet{stach19}.

We find that 17/41 SMGs are formally resolved at $>3\sigma$ significance (i.e. $\theta /\delta\theta \geq 3$) in our {\sc casa} {\sc imfit} measurements at 6\,GHz. However, deconvolved size measurements for compact sources (i.e.\ close to the beam size) must be interpreted with caution, as they may be susceptible to spurious source-broadening due to correlated noise in the image and/or residual calibration errors. As a result, at low S/N even point sources may be spuriously fit as extended sources. To account for this effect, we ran a suite of simulations using the {\sc casa} task {\sc simobserve} for 10,000 point source models with $uv$ coverage similar to that of our real 6\,GHz observations, created maps and catalogues from these simulations, and then, following \citet{bondi08}, fit an envelope of the form $S_{\rm TOT}/S_{\rm PEAK} = 1+A(S/N)^B$ to the resulting source catalogue, where $S_{\rm TOT}$ is the total flux density of a source and $S_{\rm PEAK}$ is its peak flux density. For point sources, $S_{\rm TOT}=S_{\rm PEAK}$ whereas for resolved sources $S_{\rm TOT}>S_{\rm PEAK}$. We find that the coefficients $A=2.7$ and $B=-9.8$ define an upper-limit above which fewer than $1\%$ of our point source models are artificially scattered due to noise (see Appendix A.2 for details), and hence apply this envelope as a quality control on the real data. 15/17 SMGs with measured angular sizes in Table\,1 lie above this envelope, and have a median angular size of $\theta_{\rm 6\,GHz}=0.53\pm0.07''$. Of these 15 SMGs with reliable sizes at 6\,GHz, nine are also spatially-resolved (at $>3\sigma$) in our $870\,\mu$m dust continuum maps. The median {\sc fwhm} sizes of these nine SMGs are $\theta_{\rm 6\,GHz}=0.58\pm0.10''$ and $\theta_{870\,\mu{\rm m}}=0.49\pm0.06''$, a modest factor of $\sim 1.19\pm 0.25$ (i.e.\ $\sim1\sigma$) difference, corresponding to linear scales of $4.9\pm0.8$\,kpc and $4.1\pm0.4$\,kpc, respectively. These 6\,GHz sizes are comparable to the radio sizes measured recently at 3\,GHz by \citet{miettinen17} in their study of SMGs in the COSMOS field (selected across a similar redshift range). 

At 1.4\,GHz, we find that 16/41 SMGs are reported as marginally resolved by {\sc imfit} (with sizes measured to $>3\sigma$ significance), with a median deconvolved angular size of $\theta_{\rm 1.4\,GHz}=1.35\pm0.08''$, corresponding to a linear scale of $10.9\pm0.1$\,kpc. Of these 16 SMGs, five also have both 6\,GHz and $870\,\mu$m sizes -- the median 1.4\,GHz deconvolved size of this sub-sample is $\theta_{{\rm 1.4\,GHz}}=1.35\pm0.20''$, $\sim 2.5\pm 0.4$ times their $870\,\mu$m sizes.

We report these size measurements in Table\,\ref{tab:properties}. At each observing frequency, there are a number of SMGs for which we cannot measure reliable deconvolved sizes from {\sc imfit}, which likely includes a combination of unresolved sources and sources for which a robust Gaussian fit cannot be obtained due to low signal-to-noise. To mitigate this bias, and to better understand the typical dust and radio continuum sizes of our 6\,GHz-selected SMG sample, we also measure sizes in each of the stacked subsamples presented in \S\,\ref{sect:stacking}. We find that the stacks of all 41 6\,GHz-detected SMGs have deconvolved {\sc fwhm} sizes of $\theta_{870\,\mu{\rm m}}=0.28\pm0.06''$, $\theta_{\rm 6\,GHz}=0.54\pm0.04''$ ($\theta_{\rm 6\,GHz}/\theta_{870\,\mu{\rm m}}\sim 1.8\pm 0.4$) and $\theta_{\rm 1.4\,GHz}=1.36\pm0.16''$ ($\theta_{\rm 1.4\,GHz}/\theta_{\rm 6\,GHz}\sim 2.6\pm0.4$). We also report the deconvolved sizes of the ``bright'', ``faint'' and ``convex'' stacks in Table\,\ref{tab:properties}.

\begin{table*}
\centering
\caption[The Far-IR/radio correlation and cosmic ray electron diffusion]{The Far-IR/radio correlation and cosmic ray electron diffusion}
\label{tab:diffusion}
\begin{tabular}{lcccccccccccc}
\hline
\multicolumn{1}{l}{ID} &
\multicolumn{1}{c}{$\alpha_{\rm 1.4\,GHz}^{\rm 6\,GHz}$$^a$} &
\multicolumn{1}{c}{$\alpha_{\rm 610\,MHz}^{\rm 1.4\,GHz}$$^b$} &
\multicolumn{1}{c}{$q_{\rm IR}^c$} &
\multicolumn{1}{c}{$n_{\rm H}$} & 
\multicolumn{1}{c}{$B$} &
\multicolumn{1}{c}{$\tau_{\rm cool, min}$} &
\multicolumn{1}{c}{$\tau_{\rm cool, max}$} &
\multicolumn{1}{c}{$l_{\rm cool, min}$} &
\multicolumn{1}{c}{$l_{\rm cool, max}$} &\\
\multicolumn{1}{l}{} &
\multicolumn{1}{c}{} &
\multicolumn{1}{c}{} &
\multicolumn{1}{c}{} &
\multicolumn{1}{c}{(${\rm cm}^{-3}$)} &
\multicolumn{1}{c}{($\mu$G)} &
\multicolumn{1}{c}{($\times 10^4\,{\rm yr}$)} &
\multicolumn{1}{c}{($\times 10^4\,{\rm yr}$)} &
\multicolumn{1}{c}{(${\rm pc}$)} &
\multicolumn{1}{c}{(${\rm pc}$)} &\\
\hline
AS2UDS002.1$^\star$&$-0.87\pm0.17$&$>-1.99$&$2.7\pm0.5$&$15\pm3$&$40\pm5$&$3\pm1$&$12\pm5$&$120\pm30$&$230\pm50$\\
AS2UDS003.0$^\star$&$-0.89\pm0.16$&$>-1.68$&$2.1\pm0.5$&$23\pm7$&$50\pm5$&$3\pm2$&$19\pm13$&$120\pm40$&$290\pm100$\\
AS2UDS006.1$^\star$&$-0.07\pm0.16$&$>-1.41$&$2.1\pm0.4$&$4\pm1$&$20\pm5$&$23\pm11$&$33\pm15$&$350\pm80$&$410\pm100$\\
AS2UDS013.0{\color{gray}$^\blacksquare$}&$-1.01\pm0.08$&$-0.40\pm0.22$&$2.5\pm0.2$&$13\pm1$&$35\pm5$&$5\pm1$&$18\pm2$&$150\pm10$&$270\pm20$\\
AS2UDS013.1&$-1.07\pm0.17$&$>-1.67$&$1.9\pm0.3$&$3\pm1$&$20\pm5$&$30\pm8$&$90\pm26$&$380\pm60$&$670\pm100$\\
AS2UDS015.0&$-1.06\pm0.13$&$>-2.05$&$1.5\pm0.4$&$11\pm4$&$35\pm5$&$3\pm2$&$17\pm14$&$120\pm50$&$300\pm130$\\
AS2UDS017.0&$-1.09\pm0.13$&$>-1.95$&$2.1\pm0.2$&$9\pm1$&$30\pm5$&$8\pm1$&$30\pm5$&$190\pm20$&$370\pm30$\\
AS2UDS017.1&$-1.36\pm0.10$&$>-0.67$&$2.2\pm0.4$&$4\pm1$&$20\pm5$&$20\pm10$&$68\pm34$&$300\pm80$&$550\pm140$\\
AS2UDS021.0{\color{gray}$^\blacksquare$}&$-1.18\pm0.04$&$-0.83\pm0.06$&$1.7\pm0.1$&$12\pm4$&$35\pm5$&$8\pm3$&$17\pm7$&$180\pm40$&$270\pm60$\\
AS2UDS023.0{\color{gray}$^\blacksquare$}&$-1.22\pm0.10$&$-1.13\pm0.23$&$1.9\pm0.2$&$19\pm2$&$45\pm5$&$13\pm4$&$33\pm9$&$230\pm30$&$360\pm50$\\
AS2UDS039.0&$-1.02\pm0.17$&$>-2.21$&$2.5\pm0.2$&$12\pm2$&$35\pm5$&$6\pm1$&$22\pm5$&$160\pm20$&$310\pm30$\\
AS2UDS056.1$^\star${\color{gray}$^\blacksquare$}&$-0.66\pm0.06$&$-0.04\pm0.19$&$1.9\pm0.1$&$4\pm1$&$20\pm5$&$6\pm0$&$21\pm1$&$170\pm0$&$330\pm10$\\
AS2UDS064.0$^\star${\color{gray}$^\blacksquare$}&$-1.12\pm0.03$&$-0.76\pm0.08$&$0.9\pm0.1$&$27\pm6$&$50\pm5$&$8\pm4$&$20\pm9$&$180\pm40$&$290\pm70$\\
AS2UDS072.0$^\star$&$-1.42\pm0.21$&$>-0.87$&$2.1\pm0.2$&$20\pm2$&$45\pm5$&$8\pm2$&$24\pm5$&$190\pm20$&$320\pm30$\\
AS2UDS082.0&$-0.71\pm0.19$&$>-2.42$&$2.5\pm0.3$&$14\pm2$&$35\pm5$&$21\pm6$&$33\pm9$&$300\pm40$&$370\pm50$\\
AS2UDS113.0$^\star$&$>-0.13$&$-$&$-$&$21\pm3$&$45\pm5$&$11\pm4$&$30\pm11$&$210\pm40$&$350\pm70$\\
AS2UDS116.0&$-0.99\pm0.19$&$>-0.83$&$2.3\pm0.2$&$16\pm1$&$40\pm5$&$11\pm1$&$30\pm2$&$210\pm10$&$350\pm10$\\
AS2UDS125.0&$-0.92\pm0.09$&$>-0.46$&$2.4\pm0.4$&$18\pm4$&$45\pm5$&$15\pm7$&$36\pm18$&$240\pm60$&$380\pm100$\\
AS2UDS129.0&$>-0.71$&$-$&$-$&$11\pm3$&$35\pm5$&$10\pm5$&$29\pm14$&$210\pm50$&$360\pm90$\\
AS2UDS137.0&$-0.97\pm0.07$&$-1.27\pm0.24$&$2.0\pm0.3$&$9\pm1$&$30\pm5$&$6\pm2$&$22\pm8$&$170\pm30$&$310\pm60$\\
AS2UDS238.0&$0.08\pm0.22$&$>-2.34$&$2.4\pm0.3$&$14\pm3$&$40\pm5$&$10\pm4$&$47\pm17$&$210\pm40$&$440\pm80$\\
AS2UDS259.0&$-0.89\pm0.14$&$>-0.20$&$2.5\pm0.1$&$15\pm3$&$40\pm5$&$8\pm2$&$38\pm9$&$180\pm20$&$390\pm50$\\
AS2UDS265.0&$-0.98\pm0.17$&$>-2.03$&$2.2\pm0.2$&$14\pm2$&$35\pm5$&$20\pm4$&$47\pm10$&$290\pm30$&$450\pm50$\\
AS2UDS266.0&$-1.18\pm0.14$&$>-1.54$&$2.1\pm0.3$&$15\pm4$&$40\pm5$&$19\pm8$&$43\pm20$&$280\pm70$&$430\pm100$\\
AS2UDS272.0{\color{gray}$^\blacksquare$}&$-0.73\pm0.04$&$0.21\pm0.25$&$2.4\pm0.4$&$20\pm5$&$45\pm5$&$14\pm7$&$35\pm17$&$230\pm60$&$370\pm90$\\
AS2UDS283.0{\color{gray}$^\blacksquare$}&$-1.10\pm0.19$&$-1.04\pm0.31$&$2.0\pm0.2$&$12\pm2$&$35\pm5$&$39\pm12$&$44\pm13$&$400\pm60$&$430\pm70$\\
AS2UDS297.0{\color{gray}$^\blacksquare$}&$-0.93\pm0.12$&$-0.80\pm0.28$&$2.2\pm0.1$&$12\pm1$&$35\pm5$&$19\pm2$&$50\pm6$&$280\pm20$&$450\pm30$\\
AS2UDS305.0&$-0.58\pm0.20$&$>-1.98$&$2.4\pm0.5$&$12\pm2$&$35\pm5$&$4\pm2$&$33\pm15$&$140\pm30$&$380\pm90$\\
AS2UDS311.0&$-0.69\pm0.21$&$>-1.11$&$2.3\pm0.3$&$18\pm2$&$40\pm5$&$21\pm5$&$39\pm9$&$290\pm30$&$400\pm40$\\
AS2UDS407.0{\color{gray}$^\blacktriangle$}&$-0.89\pm0.13$&$>-1.88$&$2.1\pm0.3$&$10\pm2$&$30\pm5$&$24\pm9$&$59\pm23$&$320\pm60$&$500\pm100$\\
AS2UDS412.0&$-0.40\pm0.19$&$>-1.78$&$2.4\pm0.3$&$11\pm2$&$35\pm5$&$14\pm4$&$41\pm13$&$250\pm40$&$430\pm70$\\
AS2UDS428.0{\color{gray}$^\blacksquare$}&$-0.98\pm0.05$&$-0.39\pm0.13$&$2.0\pm0.1$&$15\pm1$&$40\pm5$&$20\pm1$&$40\pm2$&$280\pm10$&$390\pm10$\\
AS2UDS460.1{\color{gray}$^\blacksquare$}&$-1.32\pm0.11$&$-1.26\pm0.26$&$1.0\pm0.1$&$6\pm3$&$25\pm5$&$35\pm28$&$88\pm69$&$410\pm170$&$650\pm270$\\
AS2UDS483.0&$-1.40\pm0.16$&$>-1.33$&$1.8\pm0.3$&$12\pm2$&$35\pm5$&$28\pm12$&$60\pm26$&$340\pm70$&$490\pm110$\\
AS2UDS497.0&$-1.04\pm0.06$&$-1.06\pm0.22$&$2.5\pm0.1$&$7\pm1$&$25\pm5$&$24\pm2$&$103\pm7$&$300\pm10$&$630\pm20$\\
AS2UDS550.0$^\star$&$-0.65\pm0.13$&$>-1.75$&$2.4\pm0.2$&$9\pm2$&$30\pm5$&$8\pm2$&$20\pm5$&$200\pm20$&$310\pm40$\\
AS2UDS590.0&$>-0.75$&$-$&$-$&$15\pm2$&$40\pm5$&$31\pm9$&$61\pm18$&$360\pm50$&$510\pm70$\\
AS2UDS608.0$^\star${\color{gray}$^\blacktriangle$}&$-0.73\pm0.14$&$>-0.45$&$2.1\pm0.1$&$14\pm8$&$40\pm10$&$12\pm11$&$33\pm29$&$220\pm100$&$370\pm170$\\
AS2UDS648.0&$-0.66\pm0.14$&$>-1.94$&$2.2\pm0.2$&$4\pm2$&$20\pm5$&$19\pm13$&$60\pm40$&$310\pm110$&$540\pm190$\\
AS2UDS665.0&$-0.73\pm0.17$&$>-2.21$&$2.4\pm0.4$&$6\pm1$&$25\pm5$&$24\pm9$&$69\pm25$&$330\pm60$&$560\pm100$\\
AS2UDS707.0$^\star${\color{gray}$^\blacktriangle$}&$-0.22\pm0.18$&$>-2.38$&$2.7\pm0.3$&$12\pm2$&$35\pm5$&$13\pm4$&$36\pm10$&$230\pm30$&$390\pm60$\\
\hline
Stack (All)&$-0.98\pm0.07$&$-0.81\pm0.11$&$2.20\pm0.13$&$10\pm1$&$35\pm1$&$4.2\pm0.6$&$35.9\pm5.5$&$130\pm10$&$390\pm30$\\
Stack (Bright)&$-1.35\pm0.24$&$-0.79\pm0.07$&$2.14\pm0.15$&$10\pm1$&$35\pm2$&$2.8\pm0.6$&$35.8\pm7.3$&$110\pm10$&$390\pm40$\\
Stack (Faint)&$-0.81\pm0.11$&$-0.77\pm0.16$&$2.49\pm0.23$&$10\pm1$&$35\pm2$&$5.8\pm1.4$&$35.6\pm8.4$&$160\pm20$&$390\pm50$\\
Stack (Convex)&$-1.06\pm0.04$&$-0.60\pm0.06$&$2.20\pm0.27$&$10\pm2$&$35\pm3$&$2.7\pm0.9$&$35.3\pm11.5$&$110\pm20$&$380\pm60$\\
\hline
\end{tabular}
\\ {\small Notes: $^{a,b}$Spectral index limits between two frequencies with one detection are determined by setting the flux density of the non-detection to $3\sigma$. It is not possible to constrain the spectral index between two non-detections. $^c$Where $\alpha^{\rm 1.4\,GHz}_{\rm 610\,MHz}$ is constrained by a 1.4\,GHz detection and a 610\,MHz upper-limit, we measure $q_{\rm IR}$ by assuming a canonical $\alpha^{\rm 1.4\,GHz}_{\rm 610\,MHz}=-0.8$, providing this assumption is consistent with the $3\sigma$ flux density limits. $^\star$,$^{\color{gray}\blacktriangle}$,$^{\color{gray}\blacksquare}$ have the same meaning as in Table\,\ref{tab:properties}}
\end{table*}

\section{Discussion}\label{sect:discussion}
\subsection{Modelling the radio spectra of SMGs}\label{sec:sed_discussion}
                      
A significant finding of our work is that a sub-sample (10/41) of our 6\,GHz detected SMGs exhibit radio spectra which steepen at higher frequency, in contrast with local ULIRGs \citep{condon92} and $\sim\mu$Jy high-redshift star-forming galaxies \citep{murphy17} which exhibit flattening spectra towards higher frequency. We now investigate this phenomenon within the context of a model that takes into account cooling timescales for cosmic ray electrons (CREs). In general, the cooling timescale of CREs at an energy $E=h\nu_{\rm C}$ is $\tau_{\rm cool}^{-1}=\tau_{\rm IC}^{-1}+\tau_{\rm sync}^{-1}+\tau_{\rm brem}^{-1}+\tau_{\rm ion}^{-1}$, with energy losses due to Inverse Compton, synchrotron, bremsstrahlung and ionization processes, respectively. In each of these processes, higher-energy electrons (which produce higher-frequency synchrotron emission) lose their energy more rapidly than lower-energy electrons (whose emission dominates the synchrotron spectrum at lower-frequencies), such that, over time, the ageing radio spectrum builds up a ``break'' at frequency $\nu_{C}$ (known as the ``critical frequency''), which moves to successively lower frequency as the CRE population ages \citep{carilli96}. \citet{thompson06} and \citet{murphy08} give approximate forms for these cooling timescales as a function of the critical frequency and properties of the host galaxy: 

\begin{center}
  \begin{equation}
    \biggl(\frac{\tau_{\rm IC}}{\rm yr}\biggr)\sim5.7\times10^{7}\biggl(\frac{\nu_{\rm C}}{\rm GHz}\biggr)^{-1/2}\biggl(\frac{B}{\mu{\rm G}}\biggr)^{1/2}\biggl(\frac{10^{-12}{\,\rm erg\,cm}^{-3}}{U_{\rm rad}}\biggr)
  \end{equation}
  \begin{equation}
    \biggl(\frac{\tau_{\rm sync}}{\rm yr}\biggr)\sim1.4\times10^{9}\biggl(\frac{\nu_{\rm C}}{\rm GHz}\biggr)^{-1/2}\biggl(\frac{B}{\mu{\rm G}}\biggr)^{-3/2}
  \end{equation}
  \begin{equation}
    \biggl(\frac{\tau_{\rm ion}}{\rm yr}\biggr)\sim10^9\biggl(\frac{n_{\rm H}}{\rm cm^{-3}}\biggr)^{-1}\biggr(\frac{\nu_{\rm C}}{\rm GHz}\biggr)^{1/2}\biggl(\frac{B}{\mu{\rm G}}\biggr)^{-1/2}
  \end{equation}
  \begin{equation}
    \biggl(\frac{\tau_{\rm brem}}{\rm yr}\biggr)\sim3\times10^{7}\biggl(\frac{n_{\rm H}}{\rm cm^{-3}}\biggr)^{-1}
  \end{equation}
\end{center}

We estimate the typical radiation field strengths of our SMGs, $U_{\rm rad}$, from their rest-frame $8$--$1000$\,$\mu$m luminosities ($L_{\rm IR}$) measured in \S\,\ref{sect:qir} via the relation $U_{\rm rad}\propto L_{\rm IR}/(2\pi R^2 c)$ for a disk radius $R$, where $c$ is the speed of light.

The magnetic field strength, $B$, is estimated under the assumption of magnetic flux freezing as used by \citet{miettinen17}, i.e.\ $B\approx 10\,\mu{\rm G}\times\sqrt{n_{\rm H}/{\rm cm}^{-3}}$. $n_{\rm H}$ is the hydrogen gas density, which we estimate from the dust masses obtained via far-IR SED fitting (\S\,\ref{sect:qir}), and using an integrated gas-to-dust ratio $\delta_{\rm GDR}=100$ that is appropriate for SMGs \citep{swinbank14}. We begin with the assumption that the majority of the cold gas in our SMG sample is located in disks whose radii are comparable to the typical \co\jonezero\ sizes of SMGs, i.e.\ $R\sim 8$--$10$\,kpc \citep{ivison11, thomson12}, with a putative vertical scale height $h\sim 1$\,kpc. The typical gas density is therefore approximately $n_{\rm H}= (M_{\rm dust}\times\delta_{\rm GDR}) / (\pi R^2 h)\sim 12\pm 2$\,cm$^{-3}$, yielding $B\sim 35\pm 3\,\mu$G\footnote{Here, the uncertainty on the magnetic field strength is purely statistical, arising from the uncertainty on the dust mass, and does not account for the inherent systematic uncertainties entailed in the choice of dust-to-gas ratio, gas disk volume, or the assumption of flux-freezing.}.

The typical radiation field strength within the region traced by the dust is $U_{\rm rad}\propto L_{\rm IR}/R_{\rm dust}^2\sim (1.5\pm0.4)\times10^{-9}$\,erg\,s$^{-1}$\,cm$^{-2}$. However, under the flux-freezing assumption discussed above, the magnetic field -- which defines the ``bath'' in which CREs lose their energy -- is traced by the gas disk, with a radius $R_{\rm gas}$ which may be up to $5\times$ larger than that of the dust disk. Throughout this larger region, the average radiation field strength is an order of magnitude weaker, $U_{\rm rad}\propto L_{\rm IR}/R_{\rm gas}^2\sim (1.2\pm0.3)\times 10^{-10}$\,erg\,s$^{-1}$\,cm$^{-2}$. 

Together, these two extreme estimates of the radiation field strength imply cooling timescales $\tau_{\rm cool}\sim 10^4$--$10^5$\,yr for the emission probed by our observed-frame 6\,GHz observations. At the relatively modest $B$-field strengths implied by the flux-freezing assumption, this timescale is determined by dominant synchrotron losses. These cooling timescales are similar to those estimated by \citet{miettinen17} for their sample of SMGs in COSMOS. 

The unexpected steepening of the radio spectra at higher frequency seen in a subset of our sample (Fig.\,\ref{fig:bright13}) implies either a severe steepening of the synchrotron emission at higher frequencies (an effect which is then mitigated by the addition of a strong, flatter-spectrum free-free component), or that the free-free component is suppressed (or absent), in which case the observed spectral curvature can be explained by more modest synchrotron steepening.

We explore these possibilities by constructing a model for the evolution of the synchrotron spectra of our galaxies. In nearby radio galaxies, there is a well-established relationship between the steepness of the radio spectrum and the age of the radio emission. If synchrotron emission is injected in to the ISM via an instantaneous event (e.g.\ a single Type II supernova), with a power-law injection index $\alpha_{\rm inj}=-0.8$ out to infinite frequency, then as the spectrum ages, high-energy electrons will, as previously discussed, lose energy (via a combination of Inverse Compton, synchrotron, ionization and bremsstrahlung processes) more rapidly than low-energy electrons, resulting in losses of radio power that are more severe at higher-frequencies. This naturally produces a steepening of the radio spectrum above a critical frequency, $\nu_{\rm C}$, and is the means by which the ages of synchrotron jets in powerful AGN are determined \citep[e.g.\ ][]{carilli96}.  For synchrotron losses, after a time $t_{\rm sync}$, the critical frequency is:

\begin{center}
  \begin{equation}\label{eq:bfieldage}
    \biggl(\frac{\nu_C}{\rm GHz}\biggr) = 1610^2\biggl(\frac{B}{\mu G}\biggr)^{-3}\biggl(\frac{t_{\rm sync}}{\rm Myr}\biggr)^{-2}
  \end{equation}
\end{center}

Assuming no pitch-angle scattering of relativistic particles, then for $\nu<\nu_C$, the low-frequency spectral index, $\alpha_{\rm L}$, remains unchanged from the injection spectral index, while at $\nu>\nu_C$, the high-frequency spectral index, $\alpha_{\rm H}$ is steepened to $\alpha_{\rm H}=(4/3)\alpha_{\rm L}-1$ \citep[the Kardashev-Pacholczyk model;][]{kardashev62,pacholczyk70}. Thus, for an instantaneous injection of synchrotron emitting electrons observed at time $t_{\rm sync}$\,(Myr), the low-frequency spectral index is expected to remain the same as the original injection spectrum ($\alpha_{\rm L}=-0.8$), while the spectral index at frequencies higher than the critical frequency becomes $\alpha_{\rm H}=-2.1$.

Of course, the radio spectra of star-forming galaxies are not the product of instantaneous injection events, but instead reflect the aggregate of the synchrotron emission produced throughout the star-formation history of the host galaxy (minus the aforementioned age/frequency dependent losses), plus the flatter-spectrum thermal free-free component tracing current star formation.

Clearly, detailed modelling of this interplay between ongoing synchrotron injection and ageing processes in distant starburst galaxies is beyond the scope of this work, however we can begin to investigate these processes by de-redshifting the radio spectrum of our ``convex'' sub-sample of SMGs and fitting the resulting rest-frame radio spectrum using a simple model for synchrotron losses. We begin by using the median far-IR luminosity of the convex sample to estimate a representative ${\rm SFR}_{\rm convex}=(850\pm 120)$\,M$_\odot$\,yr$^{-1}$, and convert this to an expected free-free luminosity density (at $\nu_{\rm rest}=20$\,GHz) of $L_{\rm FF}=(1.3\pm 0.1)\times 10^{23}$\,W\,Hz$^{-1}$ using the relations in \citet{murphy11}. Given a free-free spectral index of $\alpha_{\rm FF}=-0.1$, this allows us to estimate the thermal contribution to the rest-frame radio SED as a function of frequency, and subtract it to obtain a pure synchrotron spectrum.

We then model the evolution in the shape of this synchrotron spectrum throughout a 100\,Myr constant SFR episode (i.e.\ a ``top hat'' star formation history) by generating a grid of broken power laws from $\nu_{\rm rest}=0.1$--$40$\,GHz (arbitrarily normalised in flux at $100$\,MHz) using Equation \ref{eq:bfieldage} for instantaneous synchrotron ages $t_{\rm i}=0$--$200$\,Myr and magnetic fields that vary from $B=1$--$100\,\mu$G, and summing these aged ``instantaneous burst'' synchrotron spectra for an ongoing starburst event ``observed'' at times $t_{\rm obs}=(200-t_{\rm i})$\,Myr following the onset of star formation.

As $t_{\rm obs}$ increases, so does the age of the oldest synchrotron component present in the model spectrum ($t_{\rm sync}$), thus leading to a steepening of the spectral index at higher frequency. Due to Inverse Compton losses off the strong radiation field produced by the (ongoing) star formation, a fraction ($f(t)\propto{\rm SFR(t)}$) of this previously-injected/aged emission is suppressed at each time-step. As noted previously, Inverse Compton losses are also a function of frequency and $B$ field strength, however our three-band radio photometry do not provide sufficient constraints to simultaneously model two frequency and $B$-field dependent processes. We therefore impose the simplifying constraint that Inverse Compton losses (which are sub-dominant to synchrotron losses over the range of $B$-fields and $U_{\rm rad}$ estimated for our SMG sample) result in an additional frequency-independent suppression in the total radio power of $\sim 5$\%/Myr \citep[e.g.\ ][]{schleicher13}. We note that more detailed modelling of Inverse Compton losses as a function of frequency would likely change the derived synchrotron ages, but not the general behaviour of the synchrotron ageing model.

Hence in our model, at any time $t_{\rm obs}$ during an on-going episode of star formation, the synchrotron spectrum $\zeta_{\rm Sync}(\nu, t_{\rm obs})$ can be described as:

\begin{center}
  \begin{eqnarray*}
    \zeta_{\rm sync}(\nu, t_{\rm obs})=  \sum\limits_{i<{\rm obs}}^{} (1-f_i)\times \zeta_{\rm sync}(\nu, t_i)
  \end{eqnarray*}
  \begin{eqnarray}
    + \rho(t_{\rm obs})\times \zeta_{\rm sync}(t_{\rm obs})
  \end{eqnarray}
\end{center}

\noindent where $\zeta_{\rm sync}(\nu, t_i)$ are the individual synchrotron components in the ISM arising from injection at all prior time-steps, $t_i<t_{\rm obs}$, $f\equiv 0.05$ is the aforementioned damping coefficient which accounts for Inverse Compton losses off the local radiation field, and $\rho(t_{\rm obs})$ is a multiplicative factor which scales the amount of emission injected at each $t$-step relative to the star-formation history. In the simple case of a $100$\,Myr long ``top hat'' star-formation history (wherein the star-formation rate is constant from $0$--$100$\,Myr and then terminates), $\rho=1$ and $f$ is a constant\footnote{For more complicated star-formation histories, $\rho(t)\propto 1/f(t)$, since a higher(lower) SFR implies a higher(lower) radiation field strength, which implies more(less) rapid Inverse Compton losses of the pre-existing radio emission in the galaxy at time $t$.}.

Finally, the model synchrotron spectrum arising from this sum (which is arbitrarily scaled, but has a spectral shape that is uniquely determined by the combination of $t_{\rm obs}$, magnetic field strength and star-formation history) is normalised to match the observed radio fluxes in the three bands.

We emphasize that the arbitrary re-normalisation of this model to fit the data at each time-step means that it is not able to capture in detail the multitude of processes by which synchrotron emission is injected into and attenuated within the ISM, however our toy model can track the dependency of the (galaxy-integrated) radio spectral \textit{shape} as a function of age and magnetic field strength, allowing us to determine which (if any) combination of $t_{\rm sync}$ and magnetic field strength can reproduce the observed spectral break in our SMG composite SED. 

\begin{figure*}
\centerline{\psfig{file=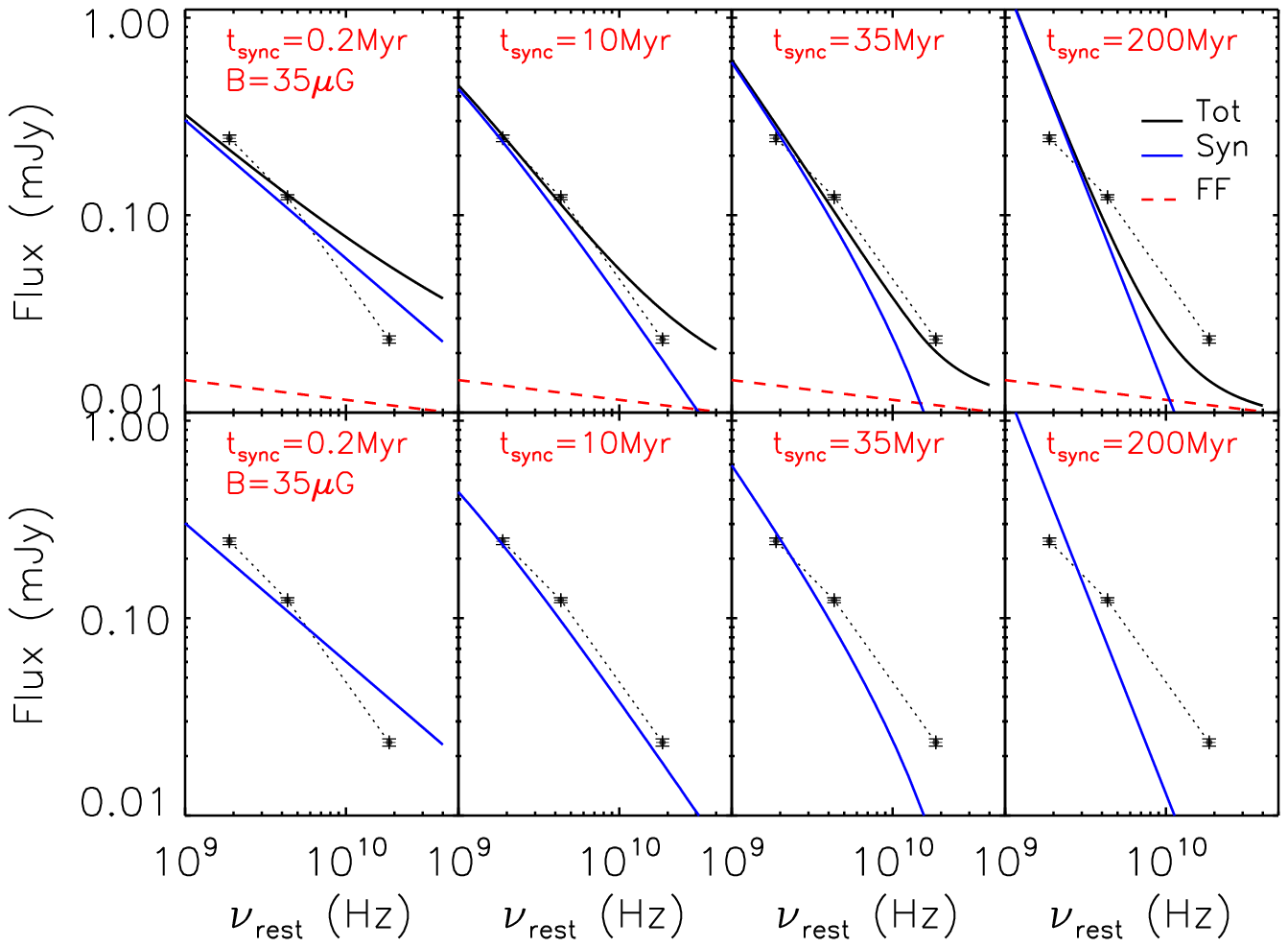,height=\textwidth,angle=90}}
\vspace*{4mm}
\caption[Short captions]{The rest-frame composite SED of our 6\,GHz-selected SMG sample, along with fitted radio spectra produced by the simple model for synchrotron ageing outlined in \S\,\ref{sec:sed_discussion}. \textit{Top row, left to right}: Example SEDs at four different ages, in which the strength of the free-free component is estimated from $L_{\rm IR}$ under the condition that ${\rm SFR}_{\rm IR} = {\rm SFR}_{\rm FF}$, and using the luminosity-to-SFR relationships of \citet{murphy11} to subtract the appropriate free-free contribution to the radio flux densities. The synchrotron model is generated as described in \S\,\ref{sec:sed_discussion}, and the total model SED is the sum of the synchrotron and free-free components. At early times ($t_{\rm sync} < 1$\,Myr), the synchrotron spectrum has a constant power-law form (with $\alpha_{\rm L} = \alpha_{\rm H} = −0.8$) from $\nu_{\rm rest}\sim 1$--$100$\,GHz which, when added to the free-free component, results in a modest flattening of the SED toward higher frequencies. As the starburst ages, a break in the synchrotron component gradually appears above a critical frequency, $\nu_{\rm C}$, as described by Equation\,\ref{eq:bfieldage}. By $t_{\rm sync} = 10$\,Myr, the ISM contains a mixture of synchrotron components with ages ${\rm 0\,Myr} < t_{\rm sync} < {\rm 10\,Myr}$ (in a proportion that is determined by the assumed star formation history), and thus the (severe) spectral steepening that has already begun to affect the oldest component(s) does not dominate the integrated SED. After the starburst terminates, there is no mechanism to inject new high-frequency emission to mitigate the (rapid) losses due to spectral ageing of previously injected components, and the spectral index is rapidly steepened to $\alpha\sim -2.1$. For a $B = 35\,\mu$G magnetic field, the optimal fit is achieved at $t_{\rm sync}\sim35$\,Myr. \textit{Bottom row, left to right}: As per the top row, except with the free-free component totally suppressed \citep[as in Arp\,220; see][and Appendix\,A.3]{barcosmunoz15}, which implies that the radio emission is dominated at all frequencies by the synchrotron component. For a given $B$-field, with no free-free component to mitigate the spectral steepening caused by the aged synchrotron component, the best-fitting synchrotron age is typically lowered by $\sim 5$--$15$\,Myr (see Appendix\,A.3)}
\vspace*{4mm}
\label{fig:bfieldspectra}
\end{figure*}

Examples of how the rest-frame SED shape evolves (for a constant star-formation rate and fixed magnetic field strength $B=35\,\mu$G, as estimated via the flux-freezing assumption) are shown in Fig.\,\ref{fig:bfieldspectra}. For the typical ISM conditions of our SMGs, we find that a synchrotron age $t_{\rm sync}\sim 35\pm10$\,Myr reproduces the observed spectral break. We stress that $t_{\rm sync}$ corresponds to the time that has elapsed since synchrotron emission first appeared in the radio spectrum of a galaxy, and is not, in general, synonymous with the age of the current starburst. We will return to this point in \S\,\ref{sect:cr-propagation}.

\subsection{The multi-frequency sizes of SMGs}\label{sect:sizesmw}

Recently, \citet{miettinen17} measured the 3\,GHz {\sc fwhm} sizes for a sample of SMGs selected at $1100\,\mu$m in the COSMOS field, finding a median $r_{\rm 3\,GHz}=4.6\pm0.4$\,kpc, a factor $\sim 1.9\pm0.2$ larger than the $870\,\mu$m dust continuum sizes measured for the AS2UDS SMGs studied here by \citet{simpson15a} and \citet{stach19}. Using estimates of the cooling times for CREs, \citet{miettinen17} argued that this apparent mis-match in the spatial scales traced by radio/sub-mm emission in SMGs cannot be due to transport of CREs produced in the dusty nuclear starburst to the outer disk region, as -- given typical ISM conditions -- the CRE cooling times are too short (by orders of magnitude) to allow propagation on the required scales.

The maximum distance ($l_{\rm cool}$) that CREs can propagate before cooling is given as $l_{\rm cool}=(D_E \tau_{\rm cool})^{1/2}$, where $D_E$ is the diffusion coefficient. Following \citet{murphy08}, we use the piecewise empirical CRE diffusion coefficient measured by \citet{dahlem95} in the local starburst galaxies, NGC\,891 and NGC\,4631:

\begin{equation} 
\label{eq-DE}
\left(\frac{D_{E}}{\rm cm^{2}~s^{-1}}\right) \sim \Bigg\{
\begin{array}{cc}
5 \times 10^{28}, & E < 1~{\rm GeV}\\
5 \times 10^{28}(\frac{E}{{\rm GeV}})^{1/2}, & E \geq 1~{\rm GeV}.
\end{array}
\end{equation}

\noindent thus 

\begin{center}
  \begin{equation}
    \biggl(\frac{l_{\rm cool}}{\rm kpc}\biggr) \sim 7\times10^{-4}\biggl(\frac{\tau_{\rm cool}}{\rm yr}\biggr)^{1/2}\biggl(\frac{\nu_{\rm C}}{\rm GHz}\biggr)^{1/8}\biggl(\frac{B}{\mu{\rm G}}\biggr)^{-1/8}
  \end{equation}
\end{center}

For the combination of $t_{\rm cool}$ and magnetic field strength estimated above, and with $D_E$ from \citet{dahlem95}, we find that CREs whose energies produce $\nu_{\rm rest}\sim 20$\,GHz radio emission have $l_{\rm cool}\sim 100$--$400$\,pc. The 6\,GHz radio {\sc fwhm} of our stacked SMGs is $\sim 1.9\pm0.4\times$ (or $2$--$3$\,kpc) larger than the $870\,\mu$m dust sizes. Given the rapid cooling timescales described above, we therefore concur with \citet{miettinen17} that diffusion of CREs from a nuclear starburst traced by the dust emission is an unlikely explanation for the enlarged radio sizes of SMGs relative to their dust sizes, unless the CRE diffusion coefficient $D_E$ in SMGs is almost three times as large as that measured empirically in local starbursts by \citet{dahlem95}. We will return to this in \S\,\ref{sect:cr-propagation}.

Comparing our new multi-frequency radio/sub-mm observations of bright SMGs to observations of ionized/molecular gas in SMGs from the literature, we see an apparent trend whereby SMGs have larger physical sizes at lower observed continuum frequencies, and that these larger low-frequency continuuum sizes successively better-trace the full extent of their diffuse ISM. To summarize, the typical {\sc fwhm} of the $870\,\mu$m (rest-frame $\sim250\,\mu$m) cold dust emission in SMGs is $\sim 0.3''$ ($r_{\rm d}\sim 2$--$3$\,kpc), while the 6\,GHz (rest-frame $\sim 20$\,GHz) radio sizes are $\sim0.5''$ ($r_{\rm 6\,GHz}\sim 4$--$6$\,kpc). At 1.4\,GHz (rest-frame $\sim 5$\,GHz), a subset of our SMGs are spatially resolved on $\sim1.3''$ ($r_{\rm 1.4\,GHz}\sim 10$\,kpc) scales, in good agreement both with previously measured 1.4\,GHz SMG sizes in the Lockman Hole field obtained with high-resolution ($\sim0.2''$) imaging from the Multi-Element Radio-Linked Interferometer Network \citep[MERLIN; ][]{biggs08}, and with the stellar disk sizes of SMGs in the ECDFS field measured via near-IR imaging with \textit{HST} \citep{aguirre13, chen15}. Moreover, observations of cold gas tracers in SMGs show the ISM of SMGs to be extended on scales of $\gtrsim 10$\,kpc \citep{ivison11, thomson12, emonts16, dannerbauer17, gullberg18}.

In a recent study, \citet{chen17} measured the \co\jthreetwo, stellar light, H$\alpha$ and $870\,\mu$m dust continuum sizes of the SMG ALESS\,67.1, finding similar size discrepancies to those quoted above \textit{within the same galaxy}, indicating that these trends and are not simply driven by biases in the individual samples used to infer them.

Throughout this paper, we have interpreted our results within the context of a model in which the ``typical'' SMG is comprised of a compact, dusty starburst -- traced by the $870\,\mu$m emission -- which acts as the source of the galaxy's primary CREs. CREs produced by a nuclear starburst are unable to propagate far from the regions in which they were injected into the ISM (due to their short life times), however cosmic ray nuclei (CRNs) may plausibly propagate outward into the more extended, quiescent gas disk \citep[e.g.\ ][]{strong98}, where they would release their energy via spallations with the baryonic content of the ISM, triggering a cascade of secondary CREs and second-generation synchrotron emission. If a large proportion of CRNs propagate and spallate in this manner, then the build-up of a low-frequency radio ``halo'' extending beyond the nuclear starburst may be expected. Because the rate of CRN/CRE production is proportional to the SFR \citep[e.g.\ ][]{papadopoulos11b}, as is the rate of dust-heating, both the total radio and total far-IR luminosities would share a common origin, and thus the far-IR/radio correlation would hold on a galaxy-averaged sense. However, any mismatch between the typical dust and \co\ spatial extents in SMGs in this model would require there to be a (local) break-down in the Schmidt-Kennicutt law, due to the implied presence of high-surface density molecular gas on the outskirts of this nuclear starburst that is not co-located with the current star-formation, traced by the submillimetre emission. We will return to this idea of CRN propagation in \S\,\ref{sect:cr-propagation}.

Alternatively, the mismatch in dust and radio continuum sizes in SMGs may be because the dust sizes themselves do not trace the full extent of the current star-formation. Dust may be driven from the central regions of a starburst galaxy either via direct photon pressure \citep[e.g.\ ][]{nath12} or by being swept-up in CR outflows \citep[e.g.\ ][]{tatischeff04, uhlig12} which propagate along magnetic field lines and terminate at large radii. Because the surface brightness sensitivity of an interferometer is inversely proportional to the angular resolution, dispersing a large fraction of dust from the nuclear starburst to a more diffuse structure could -- paradoxically -- make the dust appear more \textit{compact} in ALMA $870\,\mu$m continuum imaging, as it would shrink the region of the dust reservoir that is of sufficient surface brightness to be detectable in any given observation.

To search for signs of undetected, extended low surface brightness cold dust emission, we performed a $uv$ stack of the ALMA $870\,\mu$m continuum data for all $716$ SMGs in the UDS field. The rms of our stacked ALMA image is $\sigma_{870\,\mu{\rm m}}\sim 29\,\mu$Jy\,beam$^{-1}$, a factor $\sim 30\times$ deeper than that of a single snapshot image. In the stacked image, we do see evidence of a weak, extended dust component on $\sim 1''$ scales, however this accounts for only $\sim 1\%$ of the total $870\,\mu$m flux density. Hence we argue there is no compelling evidence for a significant mass of previously-unseen diffuse, cold dust that could enshroud a large amount of star formation, and explain the radio/sub-mm size mismatch.

To summarize, we believe the large 1.4\,GHz sizes of our SMGs (relative to their dust sizes) are most likely due to the radio emission being distributed on scales beyond those on which the bulk of the (current) star-formation is occuring, and that higher-frequency radio sizes provide a more accurate tracer of the current nuclear starburst.

\subsection{Cosmic ray propagation, and the age of the starburst}\label{sect:cr-propagation}

In \S\,\ref{sec:sed_discussion}, we determined the combinations of magnetic field strength ($B$) and synchrotron age ($t_{\rm sync}$) which best-fit the observed composite radio SED for our ``convex'' SMG sample. To convert these synchrotron ages to $t_{\rm SB}$, the time elapsed since the onset of the current starburst phase, it is necessary to consider possible time lags inherent in the model.

One contribution to the time lag, $\Delta t_{\rm SNe}$, occurs simply because the primary population of CREs, which produce synchrotron emission is built up due to supernova explosions, which occur $\sim 10$--$20$\,Myr after the onset of the starburst (i.e.\ the typical lifetimes of OB stars).

A second time lag, $\Delta t_{\rm CR}$, may occur if the transport time of cosmic rays to the regions of the ISM in which they lose their energy (i.e.\ produce their synchrotron emission) is non-negligible. Primary CRs produced in SNe comprise a mixture of relativistic antiprotons, electrons, positrons and nucleons \citep[e.g\ ][]{grenier15}, however the bulk of the synchrotron emission observed in the radio continuum is believed to originate from energy losses of short-lived CREs and positrons off the magnetic field of the host galaxy. We have already shown that barring an unusually high diffusion coefficient, propagation of primary CREs from the nuclear starburst is unlikely to explain the radio/far-IR size mismatch. However, the energy loss of primary CR nucleons (CRNs; themselves produced via supernovae) is thought to be a more complicated, multi-channel process, in which the spallations of CRNs on the baryonic content of the interstellar medium produces a cascade of secondary cosmic rays, and thus the ingredients required for secondary synchrotron emission \citep{strong07, zweibel13}.

In a study of the effect of cosmic ray streaming in hydrodynamical simulations of galaxy formation, \citet{uhlig12} found that cosmic rays produced by a compact nuclear starburst can be blown out into the ISM in a wind, which produces Alfv\'{e}n waves as it interacts with the galaxy magnetic field. While individual cosmic rays move at close to the speed of light along spiral trajectories shaped by magnetic field lines, the Alfv\'{e}n waves they produce as they do so act as a brake on the cosmic ray bulk speed, with the energy and momentum carried by the cosmic ray population being transferred to the thermal gas. This momentum transfer produces instabilities in the gas disk, which re-excite the Alfv\'{e}n waves which further limit cosmic ray streaming. Thus, the transfer of energy and momentum from cosmic rays to the interstellar medium is a self-throttling process.

In their multi-frequency radio study of the nearby starburst galaxy NGC\,253, \citet{heesen09a} approximated the typical cosmic ray bulk speed as $v_{\rm CR}\sim v_{\rm W}+v_{\rm A}$, where $v_{\rm A}=B/\sqrt{4\pi\rho}$ is the Alfv\'{e}n speed and $v_{\rm W}$ is the observed outflow speed of the thermal gas. Without spectrally and spatially-resolved observations of the molecular ISM in our SMG sample, we cannot determine $v_{\rm W}$ directly; however work by \citet{banerji11}, \citet{george14} and \citet{riechers14} has found representative outflow velocities (via [O{\sc ii}], OH and [C{\sc ii}] line emission, respectively) of  $v_{\rm W}\sim 100$--$500$\,km\,s$^{-1}$ in starbursting SMGs. Given the estimated magnetic field strengths and gas densities for our SMG sample (Table\,\ref{tab:diffusion}), we estimate a typical $v_{\rm A}\sim 1200$\,km\,s$^{-1}$, and hence $v_{\rm CR}\sim 1500$\,km\,s$^{-1}$. 

\begin{figure}
  \centerline{\psfig{file=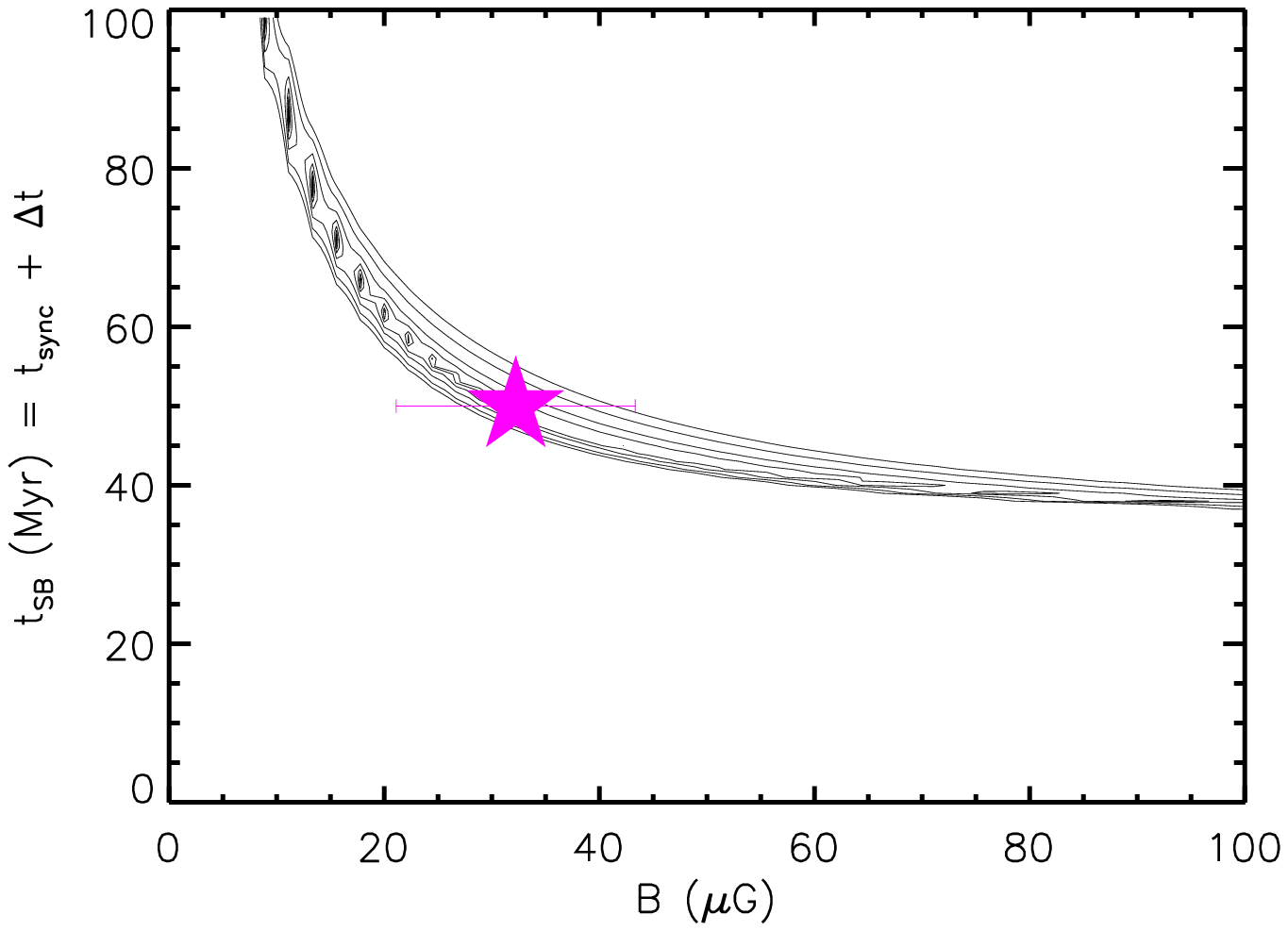,height=\columnwidth,angle=90}}
  \caption[Short captions]{$\chi^2$ contours for our model radio SEDs as a function of $B$ and $t$. The contours begin at $\chi^2=\chi^{2}_{\rm min}+1$, and increase in steps of $\chi^2_n=2\times\chi^2_{\rm n-1}$ thereafter. The statistical uncertainty on the median magnetic field strength of our SMG sample is $\Delta B\sim 2\,\mu$G (\S\,\ref{sec:sed_discussion}). As our model for the spectral shape has only one free parameter ($B$), the $\pm1\sigma$ confidence interval on the best-fitting $t$ is equal to the spacing of the innermost pair of contours at the corresponding value of $B$ (i.e.\,$\pm\Delta\chi^2=1$). The statistical uncertainties on $t$ and $B$ are thus too small to be plotted, but we reiterate that these do not account for model-dependent systematics.  Note that the ordinate is the starburst age, $t_{\rm SB}=t_{\rm sync}+\Delta t$, where $\Delta t\sim 25$\,Myr accounts for the time-lags between the onset of star-formation and the production of the observed synchrotron emission. In \S\,\ref{sect:cr-propagation}, we discuss possible contributions to $\Delta t$, including: (i) the time lag between the onset of the current starburst and the production of the first supernovae, which arises after the lifetime of a typical OB star ($\sim 10$--$20$\,Myr); (ii) the travel time of cosmic ray nucleons from the starburst to the regions of the ISM in which they lose their energy (determined by the cosmic ray wind speed) and produce synchrotron emission. Our best-estimate for the magnetic field strength is $B\sim35\,\mu$G, which implies a starburst age $t_{\rm SB}\sim 50\pm10$\,Myr.}
\vspace*{4mm}
\label{fig:chisq}
\end{figure}

If propagation and spallation of CRNs from the central, dusty starburst is the primary explanation for the $\sim 5$--$8$\,kpc mismatch in scale between the dust and 1.4\,GHz radio continuum sizes, then the implied CR propagation timescale is $\Delta t_{\rm CR}\sim 5$\,Myr. Thus, an approximate age for the starburst can be deduced from the best-fitting synchrotron age as $t_{\rm SB}\sim t_{\rm sync}+\Delta t_{\rm SNe} + \Delta t_{\rm CR}$. We show contours of starburst age ($t_{\rm SB}$) versus magnetic field strength for our best-fitting synchrotron model with these offsets applied in Fig.\,\ref{fig:chisq}. We see that, for an estimated magnetic field strength $B=35\,\mu$G, the best-fit starburst age (including time lags) is $t_{\rm SB}\sim 50\pm10$\,Myr, but note that this magnetic field strength depends on the assumption of flux-freezing, and is proportional to the gas density ($B\propto\sqrt{n_{\rm H}}$), which we have constrained only loosely via the measured dust masses. However, even allowing for a factor $4\times$ increase(decrease) in $n_{\rm H}$ would only increase(lower) the magnetic field strength by a factor $2\times$. From Fig.\,\ref{fig:chisq}, we see that for $B\sim15$--$60\,\mu$G, $t_{\rm SB}\sim40$--$80$\,Myr, and thus our estimates of the synchrotron ages of our 6\,GHz SMG sample are not strongly dependent upon our assumptions for the gas disk size and morphology.

Based on the far-IR SED fits to the 99 SMGs observed in the ECDFS field (the ``ALESS'' sample), \citet{swinbank14} used measurements of the star-formation rate (obtained via $L_{\rm IR}$) and gas mass (obtained via the dust masses and a gas-to-dust ratio) to infer typical gas depletion timescales for SMGs of $\tau_{\rm dep}\sim 130$\,Myr \citep[see also][]{bothwell13}. This is consistent with crude estimates of the lifetimes of SMGs obtained via clustering analyses \citep[e.g.\ ][]{hickox12}. From our model we find that a spectral break strong-enough to be detectable in our radio observations ($\alpha^{\rm 1.4\,GHz}_{\rm 610\,MHz}-\alpha^{\rm 6\,GHz}_{\rm 1.4\,GHz}\gtrsim 0.3$) is seen $\sim 40$--$80$\,Myr in to the star-formation event. The fact that we see such features in $\sim 25$\% of our sample suggests that the total duration of the submillimetre-bright starburst phase is thus likely to be $\sim 100$--$150$\,Myr, with those sources which display strong spectral breaks being on average at an earlier phase in their ongoing evolution than the remainder of the sample.

\section{Conclusions}\label{sect:conclusions}

We have studied the radio and rest-frame far-IR properties of a sample of 41 6\,GHz-detected submillimetre-selected galaxies from the AS2UDS survey. Combining high-resolution ($0.3''$) ALMA $870\,\mu$m imaging with radio imaging at 6\,GHz (probing $\nu_{\rm rest}\sim 20$\,GHz) at comparable resolution from the VLA, we investigate the spectral shape and relative scales of the dust and radio emission.

\begin{itemize}
\item We find that the spectral indices of radio-bright SMGs steepen toward higher frequencies in a subset of $\sim25$\% our sample (from $\langle\alpha^{1.4\,{\rm GHz}}_{610\,{\rm GHz}}\rangle=-0.60\pm 0.06$ to $\langle\alpha^{6\,{\rm GHz}}_{1.4\,{\rm GHz}}\rangle=-1.06\pm 0.04$), defying simple models for the radio emission in star-forming galaxies which predict that radio SEDs should become successively flatter at higher frequencies due to an increasing free-free component. We have investigated the possibility that our 6\,GHz flux densities are spuriously low, but find no evidence for this -- we therefore conclude that the convex spectral behaviour seen in $\sim 25$\% of our SMGs reflects their uniquely high SFR surface densities, relative to the low-SFR, $\mu$Jy radio population at high-redshift \citep{murphy17}. Our observations suggest that either synchrotron or free-free emission (or possibly both) are suppressed at high frequencies in the extreme environments of SMGs.
  
\item We develop a simple model for the radio emission in bright SMGs in which the observed spectral curvature arises due to aged synchrotron emission in the presence of an ongoing episode of intense star-formation. We use the gas masses of our SMG sample (derived from their dust masses via a gas-to-dust ratio that is appropriate for SMGs) to infer their magnetic field strengths, from which our model predicts magnetic field strengths of $B\sim 35\,\mu$G and synchrotron ages $t_{\rm sync}\sim 35\pm10$\,Myr. Accounting for the time lags between the onset of star-formation and (i) the production of the first supernovae and (ii) the propagation of cosmic-ray nucleons to $\sim$kpc radii in the gas disk (where they release their energy via interactions with the ISM, producing secondary cosmic-ray electrons, and thus additional synchrotron emission), we find that these synchrotron ages correspond to starburst ages of $t_{\rm SB}\sim40$--$80$\,Myr. These ages are consistent with estimates of the expected lifetimes of SMGs from other observations.

\item We find that the (deconvolved) 6\,GHz radio size of our stacked SMG sample is $(\sim 1.8\pm 0.4)\times$ more extended than the stacked $870\,\mu$m dust emission ($\theta_{\rm 6\,GHz}=0.51\pm 0.05''$, cf. $\theta_{\rm 6\,GHz}=0.28\pm 0.06''$), while at least a subset of our sources are spatially resolved at 1.4\,GHz, at which they are $\sim2.5\times$ larger still ($\theta_{\rm 1.4\,GHz}=1.34\pm0.18''$). We posit that this size mismatch may be consistent with the production of low-energy secondary CREs in the gas disk, far from the nuclear starburst, due to the interaction of CRNs (produced by the starburst) and baryonic material in the circum-nuclear region.

\end{itemize}

Local starburst galaxies such as NGC\,253 and Arp\,220 have cosmic ray rates $\gtrsim 10^4\times$ higher than the Milky Way \citep[e.g.\ ][]{meijerink11}. In such extreme cosmic-ray dominated regions (CRDRs), the minimum gas temperature may be elevated by as much as $\sim 100$\,K, becoming thermally decoupled from the dust. At these temperatures, astro-chemical models of CRDRs predict the efficient transmutation of \co\ into atomic/ionized carbon, and hence of elevated [C{\sc i}] and [C{\sc ii}] to low-$J$ \co\ line ratios \citep[e.g.\ ][]{bisbas17}. Hence, an important future test of our model -- which posits that the multi-frequency radio sizes and source-integrated radio spectra of SMGs might be the resulf of transport and spallations of CRNs in the extended gas disk -- will be to perform a resolution-matched spectral line survey of SMGs in order to search for radial variations in the \co/[C{\sc i}] and \co/[C{\sc ii}] profiles of SMGs.
  
Disentangling the complex spectro-morphological properties of high-redshift starburst galaxies is challenging, given the capabilities of the current VLA (particularly, its lack of resolving power at low frequencies). However these results (both the convex source-integrated radio spectra and enlarged sizes at lower frequency relative to higher frequency) provide tantalising evidence that the processes of cosmic ray propagation -- which dominate the spectral behaviour of, and create low-frequency radio continuum halos around nearby starburst galaxies \citep[e.g.\ ][]{heesen16, mulcahy18} -- may also be at work (and be observable) in the environments of high-redshift starburst galaxies. Forthcoming instruments (such as SKA, and the proposed ngVLA) and deep imaging surveys being undertaken with existing longer-baseline interferometers \citep[such as the \emerlin\ Galaxy Evolution survey, which maps the GOODS-N field at 1.4\,GHz at $\sim 0.3''$ resolution down to $\sim\mu$Jy\,beam$^{-1}$ sensitivites; T.\,W.\,B.\ Muxlow et al., in prep][]{thomson19} will provide the capabilities to create sub-arcsecond, resolution-matched spectral index maps of high-redshift galaxies. These will allow us to directly address this issue, yielding new constraints on the mechanisms powering the radio emission in starburst galaxies at high-redshift.

\section*{Acknowledgments}
We would like to thank the anonymous reviewer for their useful comments which greatly improved the content and presentation of this paper. APT, IRS, EAC \& BG acknowledge support from the ERC Advanced Grant {\sc dustygal} (\# 321334). APT, IRS, EAC, AMS, BG and JLW acknowledge STFC (ST/P000541/1). APT is immeasurably grateful to David Rosario, Rob Ivison, Eva Schinnerer, Rob Beswick, Tom Muxlow and Todd Thompson who have all provided helpful advice and insight throughout the preparation of this manuscript. IRS also acknowledges support from a Royal Society Wolfson Merit Award. EI acknowledges partial support from FONDECYT through grant N$^\circ$\,1171710. WR is supported by the Thailand Research Fund/Office of the Higher Education Commission Grant Number MRG6080294 and Chulalongkorn University's CUniverse. MJM acknowledges the support of the National Science Centre, Poland, through the POLONEZ grant 2015/19/P/ST9/04010; this project has received funding from the European Union's Horizon 2020 research and innovation programme under the Marie Sk{\l}odowska-Curie grant agreement No. 665778. We gratefully acknowledge funding towards the VLA 3-bit samplers used in this work from ERC Advanced Grant 321302, {\sc cosmicism}. JLW acknowledges support from a European Union COFUND/Durham Junior Research Fellowship (EU grant agreement number 609412) and a STFC Ernest Rutherford Fellowship (ST/P004784/1 and ST/P004784/2). We are grateful to the staff at UKIRT and JCMT for their efforts in ensuring the success of the UDS project. 

\bibliographystyle{aasjournal.bst}
\bibliography{ref.bib}

\FloatBarrier
\appendix
\subsection{A.1 -- Postage stamp images}

\begin{figure*}
\centerline{\psfig{file=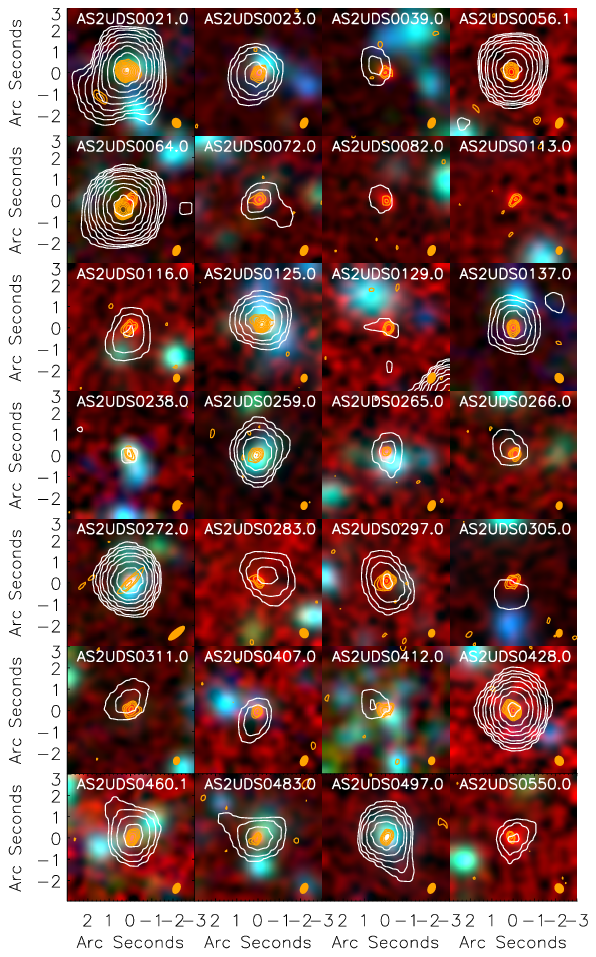,height=\textheight}}
\vspace*{4mm}
\vspace*{4mm}
\label{fig:morestamps}
\end{figure*}
\begin{figure*}
\centerline{\psfig{file=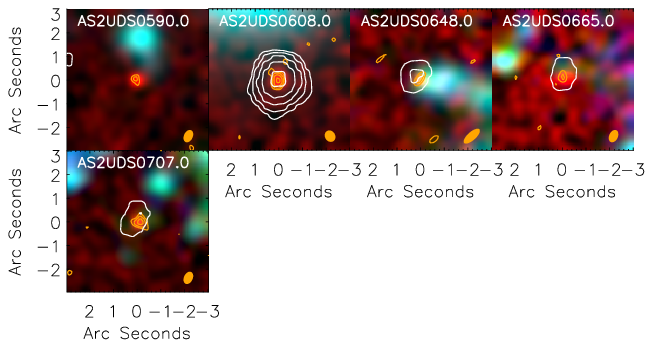,width=\textwidth}}
\vspace*{4mm}
\vspace*{4mm}
\label{fig:morestamps2}
\end{figure*}
\FloatBarrier
\vspace*{-15mm}\small{Continuation of Fig\,\ref{fig:stamps} -- false-colour thumbnails comprised of ALMA $870\,\mu$m (red), and Subaru $i$ (green) and $V$-band (blue), smoothed with a common $0.35''$ {\sc fwhm} Gaussian, with 6\,GHz (orange) and 1.4\,GHz (white) contours overlaid at $-3$, $3$, $3\sqrt{2}\times\sigma$ (and in steps of $\sqrt{2}\times\sigma$ thereafter), highlighting the morphological diversity of our sources as a function of wavelength. We show the VLA 6\,GHz synthesized beam as an orange ellipse in the bottom-right corner of each sub-figure.}

\subsection{A.2 -- Flux recovery and deconvolved size estimates in high-resolution radio maps}\label{app:tests}
\FloatBarrier

We perform a series of tests in order to confirm the reliability of both the flux densities and sizes measured from our 6\,GHz maps: first, we checked the absolute flux calibration of our reduced data by concatenating all scans of the phase calibrator source (J0215--0222) obtained under VLA project 15A-249, and making a multi-scale, multi-frequency synthesis ({\sc msmfs}) continuum image with natural weighting. The resolution of this image is $0.52'' \times 0.37''$ with a beam position angle $\theta=-41^\circ$, and we measured the flux density using the {\sc casa} {\sc imfit} tool, recovering $S_{\rm 6\,GHz}=684\pm1$\,mJy. This flux density is within $\sim 4\%$ of the canonical flux density quoted in the NRAO flux calibrator manual\footnote{\url{http://www.aoc.nrao.edu/~gtaylor/csource.html}} ($S_{\rm 6\,GHz}=710$\,mJy), indicating that the low target fluxes at 6\,GHz are unlikely to be due to a systematic error in the absolute flux calibration of our data.

\begin{figure}
  \centerline{\psfig{file=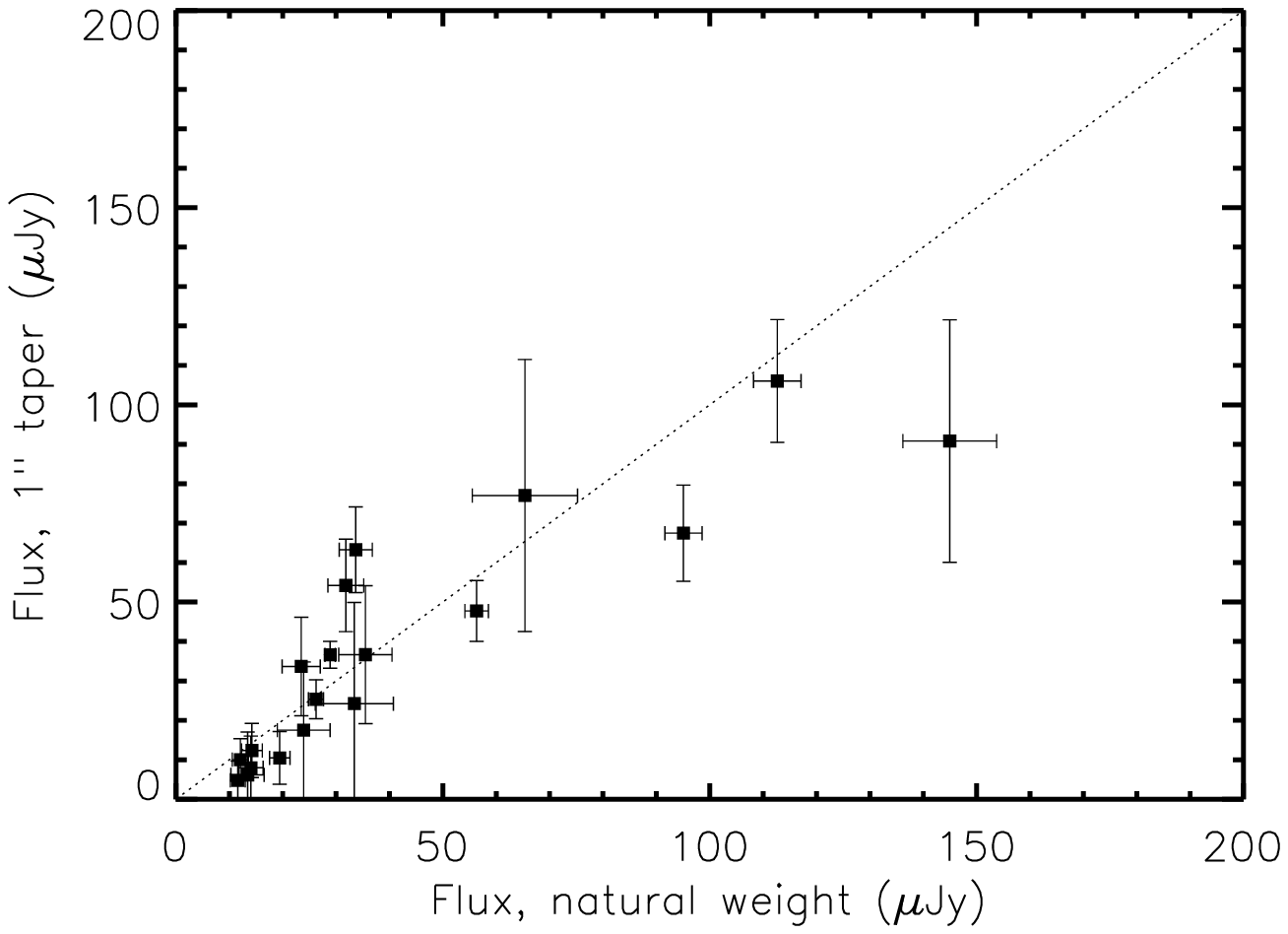,height=10cm,angle=90}}
  \caption[Short captions]{A comparison of the 6\,GHz flux densities for 21 SMGs in the naturally-weighted images used for analysis throughout this paper, with those measured in maps tapered to $1''$ resolution. If the spectral curvature seen in our 6\,GHz-selected SMG sample were the result of having resolved-out flux on $\gtrsim 1''$ spatial scales, we would have expected to have recovered more flux in the tapered maps than in the naturally-weighted maps. Instead, we find a good correspondence between the flux densities of SMGs detected in both sets of images, suggesting that our naturally-weighted 6\,GHz maps recover most of the flux in these systems, and that their low 6\,GHz flux densities are unlikely to have been be induced by the effects of $uv$ coverage.}
\vspace*{4mm}
\label{fig:flux_comparison}
\end{figure}

Next, we searched for signs that our naturally-weighted, A-configuration 6\,GHz observations may be insensitive to extended high-frequency emission by performing two independent tests. First, we re-imaged the data in {\sc wsclean} using the {\tt -taper-gaussian 1asec} option, which calculates and applies the required $uv$ taper to degrade the image-plane resolution to $1.0''$, and created updated background and rms maps via boxcar smoothing. Tapering the $uv$ data down-weights the longer baselines and increases the beam area, trading off a (modest) loss in point-source sensitivity for an increase in sensitivity to extended emission. We performed blind source extraction on the tapered images using {\sc aegean} as described in \S\,\ref{sect:obs}, and then cross-checked the flux densities of sources detected in the tapered 6\,GHz images with those of the naturally-weighted maps. We detect only 21 SMGs in the tapered 6\,GHz maps (compared to 41 SMGs in the untapered, naturally-weighted maps), with good agreement in the flux densities for sources detected in both sets of maps (Fig\,\ref{fig:flux_comparison}). We therefore find little evidence for any previously-missed extended component on $\gtrsim 1''$ scales, suggesting that most of the 6\,GHz flux is indeed located on the longer baselines which trace compact $\sim 0.5''$ structures.

As an additional check (both against resolving-out extended emission, and that our deconvolved source size estimates are probing real physical structures and are not merely the result of point sources which are broadened due to noise), we designed a suite of simulated 6\,GHz datasets using the {\sc casa} task {\sc simobserve}, and then imaged, performed object detection and measured deconvolved source sizes for these using the same workflow as was used for the real data (i.e.\ imaging the simulated datasets with {\sc wsclean}, performing object detection above a local $5\sigma$ threshold using {\sc aegean}, and then measuring the flux densities and deconvolved source sizes of detected sources with the {\sc casa} {\sc imfit} task). These simulated observations were designed to have similar $uv$ coverage to our real 6\,GHz observations, i.e.\ using the VLA A-array antenna configuration file and with the simulated observations taking place in 2\,hour chunks over a similar LST range to the real observations. Realistic projection of our source models on to the VLA baselines was ensured via the World Coordinate System information specified in the {\sc fits} header of the model image (i.e.\ using coordinates at the centre of the UDS field). We construct two classes of model image:

\subsubsection{Extended source models}

The largest angular scale ($\theta_{\rm LAS}$) defines the largest-scale structures to which an interferometer is sensitive, and can be determined theoretically from the fringe spacing formed by the shortest baseline in the array, $\theta_{\rm LAS}\sim \lambda / B_{\rm min}$. Emission from structures larger than $\theta_{\rm LAS}$ forms fringes which destructively interfere (i.e.\ are ``resolved-out''). For 6\,GHz observations undertaken with the VLA in A-configuration (where $B_{\rm min}=680$\,m), we expect $\theta_{\rm LAS}\sim 9''$ -- thus for simulated observations obtained in this array/frequency combination we expect to be able to accurately recover the flux from extended sources and for this flux to be constant up to a source {\sc fwhm} of at least several arcseconds, corresponding to $\gtrsim 60$\,kpc at the median redshift of our sample ($z\sim 2.4$). This limiting physical extent is around $\sim 8\times$ larger than the typical optical half-light radii of SMGs \citep{swinbank04, chen17}, $\sim 10\times$ larger than their millimetre-wave dust continuum sizes \citep{simpson15a, ikarashi17}, $\sim 4\times$ larger than their $1.4$\,GHz radio continuum sizes \citep{muxlow05, biggs08, jimenezandrade19} and $4\times$ larger than their maximum reported cold ISM spatial extents \citep{carilli11, thomson12, riechers13}, and therefore offers an exceedingly generous upper-limit on the scales to which our 6\,GHz observations need to be sensitive.

\begin{figure}
  \centerline{\psfig{file=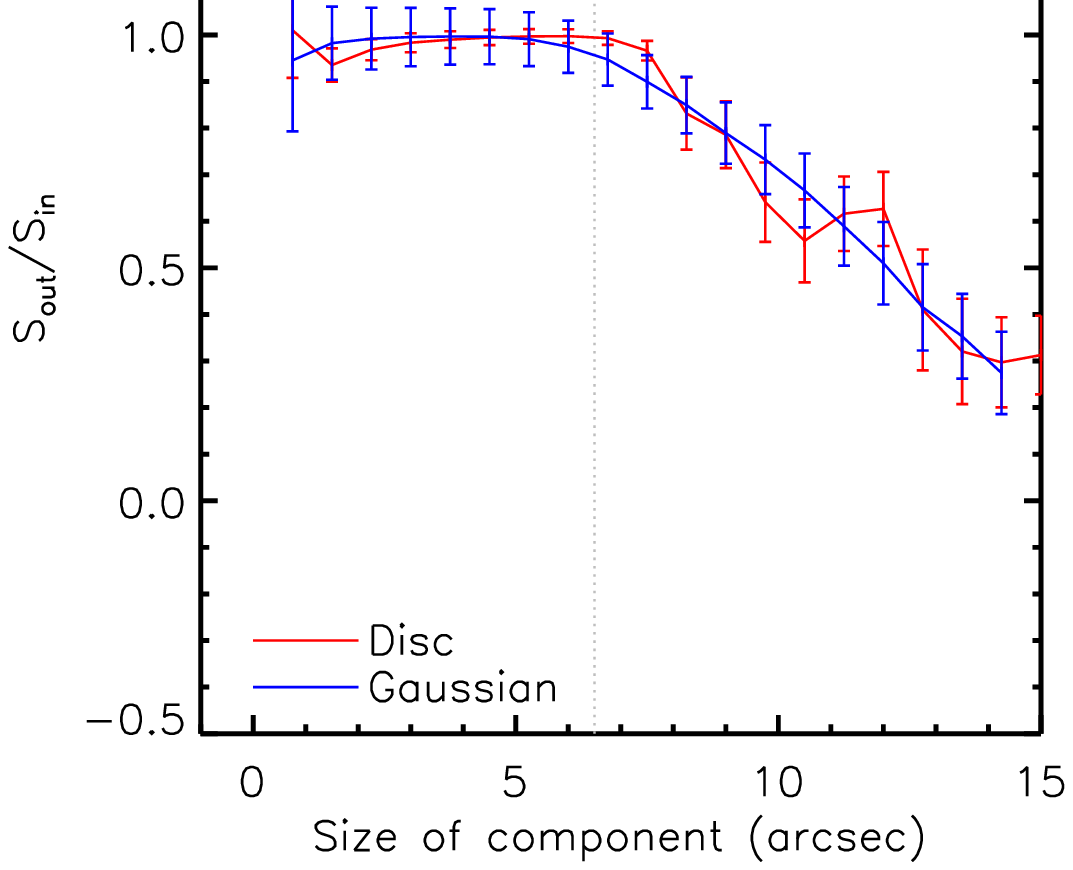,height=10cm}}
  \caption[Short captions]{Ratio of recovered versus input flux densities for simulated observations of model sources undertaken with the {\sc casa} {\sc simobserve} tool, with the array configuration, target declination and LST range chosen to provide equivalent $uv$ coverage to that of our 6\,GHz VLA programme. We see that for both Gaussian source models (with $0<${\sc fwhm}$<6.5''$) and constant surface brightness disc models (with diameter $0<D<7''$) the array can successfully recover $\sim 100\%$ of the input model flux, with a drop-off in sensitivity to flux on larger scales due to the lack of short spacings in the array. The angular scale at which this drop-off begins is very close to the theoretical largest angular scale of the array for sources observed at zenith in this array configuration ($\theta_{\rm LAS}\sim \lambda / B_{\rm min}\sim 9''$). We use coloured arrows to highlight the typical angular sizes of SMGs measured in $870\,\mu$m dust continuum imaging with ALMA \citep[$\sim 2$\,kpc; black arrow,][]{simpson15a}, resolved 1.4\,GHz \emerlin\ imaging \citep[$\sim 5$\,kpc; green arrow,][]{biggs08} and VLA observations of cold molecular gas \citep[\co\jonezero, $\sim 16$\,kpc, orange arrow;][]{ivison11}. The empirical largest angular size of the VLA in A-Array at 6\,GHz for sources observed as the UDS field transits above Socorro is $\sim 6.5''$, corresponding to $\sim 60$\,kpc at $z=2.3$. Hence it is \textit{highly} unlikely that the low observed 6\,GHz continuum flux densities of our sources are the result of resolving out significant extended emission.}
\vspace*{4mm}
\label{fig:vlalas}
\end{figure}

To confirm the estimated $\sim 9''$ largest angular scale of the array and confirm our sensitivity to extended emission on the \textit{relevant} range of spatial scales, we constructed a suite of 40 model images comprised of a single source (either a Gaussian or constant surface brightness disc model), placed at the phase centre, with sizes ({\sc fwhm} for the Gaussian model; diameter $D$ for the disc model) between $0$--$15''$. These model images were used as the input models for 40 simulated observing runs, executed with the {\sc casa} {\sc simobserve} task. We use an integration interval of 10\,s and a total observing time of 2\,hours for each simulated observation in order to provide a close match for the $uv$ coverage of our real 6\,GHz observations. The resulting simulated Measurement Sets were then imaged with {\sc wsclean}, yielding a typical synthesized beam of $\sim 0.48\times0.40''$ and a noise level $\sigma=5\,\mu$Jy\,beam$^{-1}$ with flux densities measured via aperture photometry for apertures $1.1\times$ the (known) extent of the input model. We find that for both Gaussian and constant surface brightness disc models, our chosen VLA array configuration is capable of recovering essentially all the flux of the input model for source sizes ({\sc fwhm} and diameter, respectively) up to $\sim 6.5''$, and does not begin to significantly resolve-out emission until the input source sizes are $\gtrsim 7''$. We show this drop-off in recovered flux as a function of input source size in Fig.\,\ref{fig:vlalas}, which also includes symbols marking the typical dust continuum, 1.4\,GHz and molecular gas sizes of SMGs at the same redshift. Our empirically-derived largest angular scale is in excellent agreement with the expected $\theta_{\rm LAS}$, and is several times larger than the angular sizes of SMGs as seen in any other waveband. This strongly suggests that the lower-than-expected 6\,GHz flux densities of our sources are not the result of resolving-out extended flux.

\subsubsection{Point source models}

At modest S/N ($\lesssim 10$), image artefacts (including random noise, ripples and other low-level calibration/data processing errors) may conspire to artificially broaden Gaussian fits to compact sources. To test the hypothesis that our deconvolved 6\,GHz source sizes are simply the result of spurious broadening of unresolved sources due to noise (as opposed to real extended emission) we constructed a further 5 model images comprised of $2,000$ point sources each (initially distributed randomly within the model map, but with a check to remove any sources placed within $3''$ of any other source in order to side-step issues associated with source blending), and with flux densities randomly distributed between $10$--$200\,\mu$Jy. Again, these 5 model images served as the input models for 5 successive {\sc casa} {\sc simobserve} runs, which were again imaged and catalogued following the same workflow as was used for the real 6\,GHz maps.

\begin{figure}
   \centerline{\psfig{file=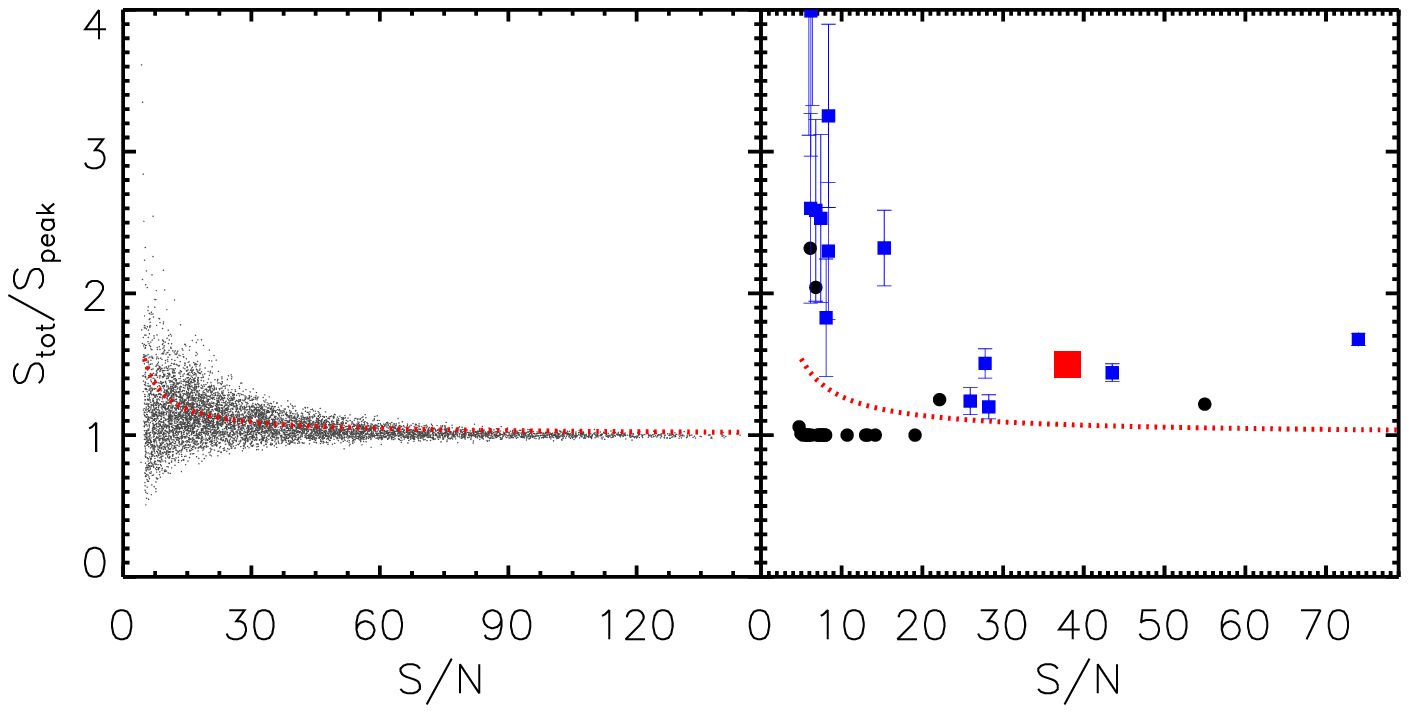,height=\textwidth,angle=90}}
  \caption[Short captions]{\textit{Left:} Ratio of integrated to peak flux densities  measured by {\sc casa} {\sc imfit} as a function of S/N for simulated VLA 6\,GHz observations of $\sim 10,000$ point sources located in the UDS field. These simulated observations were executed using the {\sc casa} {\sc simobserve} task and were designed to produce similar $uv$ coverage to our real 6\,GHz observations (i.e.\ utilising the VLA A-configuration antenna positions, and observing in 2\,hour chunks over a comparable LST range), and were subjected to an identical imaging and cataloguing workflow as was used for the real data (i.e.\ maps were made using {\sc wsclean}, which were catalogued with {\sc aegean} using a $5\sigma$ threshold for object detection before Gaussian components were fit to identified sources using the {\sc casa} {\sc imfit} routine). To quantify the impact of spurious source-broadening due to noise effects we measure the envelope (shown in red) below which $\sim 99\%$ of the sample lies as a function of S/N (see text for details), i.e.\ only $\sim 1\%$ of the point source sample is scattered above this envelope by noise. \textit{Right:} The ratio of total-to-peak flux density versus S/N for the 41 6\,GHz detected SMGs in AS2UDS. Sources which are fit as point sources (or which are rejected as extended sources because $\theta / \delta\theta < 3$) are shown as open circles, while the 15 sources which have $\theta / \delta\theta \geq 3$ (and are thus classified as ``resolved'' by {\sc casa} {\sc imfit}) are shown as blue squares. The ``convex'' stacked sub-sample is represented with a large red square, and it, along with 15 individual SMGs lie above the empirically-measured envelope which highlights the domain of plausibly up-scattered sources. Hence we classify them as being securely spatially-resolved, and report their angular sizes in Table\,\ref{tab:properties}.}
\vspace*{4mm}
\label{fig:pt_src_or_resolved}
\end{figure}

To quantify the level of spurious source-broadening it is helpful to plot the ratio of total to peak flux densities ($S_{\rm TOT}/S_{\rm PEAK}$) for recovered sources as a function of S/N for all 10,000 simulated point sources: for \textit{bona fide} point sources, $S_{\rm TOT}/S_{\rm PEAK}\sim 1$, while $S_{\rm TOT}/S_{\rm PEAK}>1$ highlights that the source-fitting procedure has determined that the source is spatially-resolved. We show $S_{\rm TOT}/S_{\rm PEAK}$ versus S/N for our point source models in Fig.\,\ref{fig:pt_src_or_resolved}. To determine the reliability of source size estimates as a function of S/N we follow the examples of \citet{bondi08} and \citet{smolcic17} and empirically determine the envelope below which 99\% of the simulated datapoints lie in bins of S/N. We fit a curve of the form $S_{\rm TOT}/S_{\rm PEAK}= 1+A(S/N)^B$ to this envelope, finding that the coefficients $A=2.7$ and $B=-9.8$ provide a good fit; only 1\% of our simulated point sources suffer from severe-enough spurious broadening to be elevated above this line, and therefore we use this empirical envelope to provide further quality assurance on the deconvolved 6\,GHz source sizes reported in \S\,\ref{sec:sizes}. Of the 17 6\,GHz-detected SMGs which are reported as ``resolved'' by {\sc casa} {\sc imfit} (i.e.\ satisfying the criteria that ${\rm S/N}\geq 5$ and $\theta / \delta\theta \geq 3$), we find that two (AS2UDS\,003.0 and 407.0) lie below this source-broadening envelope. We therefore conservatively re-classify these sources as being unresolved 6\,GHz detections. The remaining 15 SMGs -- and the stacked sub-samples -- all lie comfortably above this envelope, and are hence considered to be resolved sources.

\subsubsection{The uncertainties on deconvolved source sizes}

Finally, and returning to our real 6\,GHz data, we conduct an assessment of the reliability of the algorithm used by {\sc casa} {\sc imfit} to compute deconvolved angular source sizes ($\theta$) and their uncertainties ($\delta\theta$) from their fitted (i.e.\ image-based) size and uncertainty ($\phi$ and $\delta\phi$, respectively).

In an important work, \citet{condon97} outlines the formalism for determining errors in elliptical Gaussian fits in the presence of correlated noise. Using Equation\,21 of \citet{condon97}, we see that a source with a fitted axis size $\phi$ has an uncertainty $\sigma_{\phi}$ on that fitted size of

\begin{center}
  \begin{equation}
    \biggl( \frac{\sigma_\phi}{\phi}\biggr)^2 \approx \frac{2}{\rho^2}
    \end{equation}
\end{center}

where in Equation\,41 of the same work it is shown that

\begin{center}
  \begin{equation}
     \rho^2 = \frac{\phi_{\rm M}\phi_{\rm m}}{4\phi_{\rm N}^2}\biggl[ 1+ \biggl(\frac{\phi_{\rm N}}{\phi_{\rm M}}\biggr)^2\biggr]^{\alpha_{\rm M}}\biggl[ 1+\biggl(\frac{\phi_{\rm N}}{\phi_{\rm m}}\biggr)^2 \biggr]^{\alpha_{\rm m}}\frac{S_{\rm peak}^2}{\sigma_{\rm peak}^2}
  \end{equation}
\end{center}

\noindent is the overall signal-to-noise ratio of the Gaussian fit. Here, $\phi_{\rm M}$ and $\phi_{\rm m}$ are the fitted major/minor axis sizes, respectively, $\phi_{\rm N}$ is the angular scale ({\sc fwhm}) on which the image noise is correlated, i.e.\ approximately the synthesized beam size. The exponents $\alpha_{\rm M}/\alpha_{\rm m}$ were empirically-derived by \citet{condon97} for a variety of source models, with $\alpha_{\rm M}\sim 5/2$ and $\alpha_{\rm m}\sim 1/2$ being found for source sizes close to the telescope synthesized beam, i.e.\ marginally-resolved sources, as we have here. $S_{\rm peak}/\sigma_{\rm peak}\equiv {\rm SNR}$ is the ratio of the peak amplitude of the Gaussian fit to the local rms.

We compute uncertainties on the fitted major/minor axis sizes of our 6\,GHz resolved SMG sample using these relations, finding them to be in good agreement with the fitted size errors reported by {\sc casa} {\sc imfit}. Next, we propagate these uncertainties on the fitted source sizes through to the \textit{deconvolved} source sizes (i.e.\ the intrinsic sizes of our sources after deconvolving the telescope PSF) using Equations\,2 and 3 of \citet{murphy17}:

\begin{center}
\begin{equation}
\theta = \sqrt{\phi^2 - \theta_{\rm 1/2}^2}
\end{equation}
\begin{equation}
\biggl(\frac{\delta\theta}{\delta\phi}\biggr) = \biggl[1-\biggl(\frac{\theta_{\rm 1/2}}{\phi}\biggr)^2\biggr]^{\rm -1/2}
\end{equation}
\end{center}

\noindent where $\theta_{\rm 1/2}$ is the fitted beam size. 

\begin{figure}
   \centerline{\psfig{file=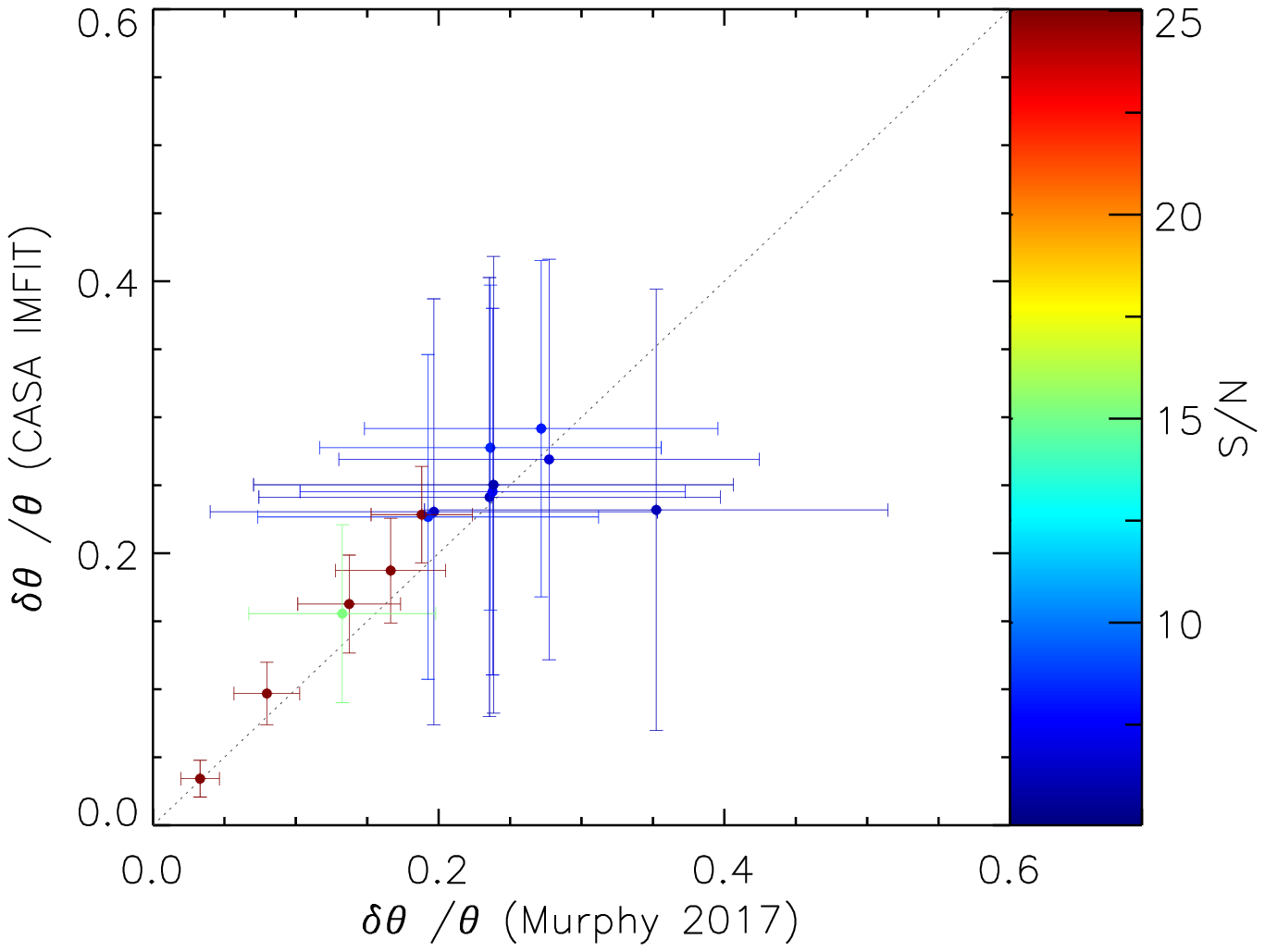,height=10cm,angle=90}}
  \caption[Short captions]{Comparison of the 6\,GHz fractional deconvolved source size error estimates (i.e.\ $\delta\theta / \theta$, where $\theta$ is the deconvolved source size and $\delta\theta$ is the error on this value) determined by the {\sc casa} {\sc imfit} routine versus those calculated from the fitted (i.e.\ non-deconvolved) source sizes/errors using the relations of \citet{condon97} and \citet{murphy17} for the 16 SMGs with measured 6\,GHz sizes in Table\,\ref{tab:properties}. Points are colour-coded by the peak S/N, and the error bars show the fractional uncertainty in the peak fluxes (i.e.\ N/S). We see excellent agreement between the deconvolved source size estimates measured by {\sc casa} {\sc imfit} and those computed using the \citet{condon97} and \citet{murphy17} relations across our sample, within the error bars set by the S/N of the data, suggesting that the deconvolved source size uncertainties reported by {\sc casa} are reliable at least down to the $5\sigma$ flux cut used for source-finding in \S\,\ref{sect:obs}.}
\vspace*{4mm}
\label{fig:size_errors}
\end{figure}

Specifically, we are interested in the fractional error on the deconvolved source sizes, $\delta\theta / \theta$, and in studying how this evolves with S/N. In general, we would expect sources detected at lower-S/N to have higher fractional uncertainties on their deconvolved sizes, however any systematic errors with the way in which {\sc casa} {\sc imfit} propagates the uncertainty on the fitted size through to the uncertainty on the deconvolved source size might be expected to produce a second-order effect which correlates with the S/N of the observation. We return to the {\sc casa} {\sc imfit} outputs for the 15 SMGs we formally classify as ``resolved'' (Table\,\ref{tab:properties}) and compare $\delta\theta_{\rm CASA}$ with $\delta\theta_{\rm Murphy, 2017}$ (which we compute from the fitted size uncertainties calculated above) in Fig.\,\ref{fig:size_errors}. We see excellent agreement between the deconvolved source sizes (and their uncertainties) returned automatically by the {\sc casa} {\sc imfit} source fitter and those calculated from the (non-deconvolved) Gaussian fits using the relations of \citet{condon97} and \citet{murphy17}.

Hence, our 6\,GHz source size estimates are most likely {\bf not} the result of spurious broadening of unresolved emission due to image-plane noise and/or due to errors in the fitting procedure used in {\sc casa} {\sc imfit}.

\subsection{A.3 -- A note on possible free-free suppression}
In a recent study on the radio spectral properties of the nuclear disks of Arp\,220, \citet{barcosmunoz15} observed a remarkably consistent spectral index $\alpha\sim-0.7$ in four frequency ranges between $4.7$--$43.5$\,GHz. Hence, the Arp\,220 radio SED neither steepens toward higher frequencies (as happens in our SMG sample) nor flattens toward higher frequency due to free-free emission (as is seen in local, ``normal'' star-forming galaxies). \citet{barcosmunoz15} argue that both this SED shape, and the high brightness temperatures ($\sim 10^4$\,K at $6$\,GHz) of Arp\,220 indicate a radio SED that is dominated at all frequencies by non-thermal synchrotron emission, with no (or little) need for any additional contribution arising from a flatter-spectrum, thermal free-free component. While the high star formation rates (${\rm SFR}\sim 200$\,M$_\odot$\,yr$^{-1}$) in the disks of Arp\,220 suggest that a strong free-free component \textit{should} be present, \citet{barcosmunoz15} argue that in sufficiently dusty environments, a non-negligible fraction of ultraviolet photons can be absorbed \textit{before} they produce the ionizations which are a prerequisite for thermal radio emission, thus lowering the thermal luminosity density relative to that expected, given the star formation rate.

The high dust masses of SMGs \citep[$M_{\rm dust}\geq 10^{8}$\,M$_\odot$; e.g.\ ][]{santini10, swinbank14} and the large optical extinctions in their nuclear regions \citep[$A_V\geq 500$; e.g.\ ][]{simpson17} certainly suggest that if such suppression of thermal emission is indeed possible, then SMGs represent an environment in which this phenomenon could be important\footnote{Alternatively, if the IMF is top-heavy -- a claim first made by \citet{baugh05}, as a necessary condition for matching observed and theoretical number counts of starburst galaxies, but recently lent its first direct observational support in the form of CNO isotope yields in starburst galaxies that are consistent with enhanced production of AGB stars \citep{romano17} -- then the resulting enhancement of $L_{\rm IR}$ per unit SFR may imply up to a factor $\sim4\times$ enhancement in the $L_{\rm IR}/{\rm SFR}$ ratio compared to that used in \citet{kennicutt98}, which assumed a \citet{salpeter55} IMF. Such an enhancement in the $L_{\rm IR}/{\rm SFR}$ ratio would naturally lead to an over-prediction of the free-free luminosity density if inferred from $L_{\rm IR}$ without taking the top-heavy nature of the IMF into account.}.

To test whether the conclusions of our model for synchrotron spectral ageing are sensitive to the assumed strength of the free-free component, we re-ran the model for $t_{\rm sync}=0$--$200$\,Myr and $B=1$--$100\,\mu$G under the extreme assumption that the free-free component is completely suppressed -- that is, requiring that the aged-synchrotron models fit the observed composite SED directly, with no prior subtraction of a scaled $\alpha\sim-0.1$ power law contribution (Fig.\,\ref{fig:bfieldspectra}). We find that suppressing the thermal component in this manner allows for synchrotron ages that are $\sim 5$--$15$\,Myr lower (for a given magnetic field strength) than are obtained from models in which the thermal component scales with the far-IR derived star formation rate, ${\rm SFR}_{\rm IR}$, but that the interpretation of our models is not fundamentally changed. We show examples SEDs for the synchrotron fits without free-free emission in Fig.\,\ref{fig:bfieldspectra}.

\subsubsection{A.4 -- AS2UDS\,0017.1: a radio-bright, gas-rich, $870\,\mu$m-faint galaxy at $z\sim 2.6$}
The 6\,GHz thumbnail of one of our sources \citep[AS2UDS\,0017.1; previously published under the ID UDS\,306.1 by][]{simpson15a, wardlow18} shows an extension/secondary component located $\sim 1''$ ($\sim 8$\,kpc) from the peak of the $870\,\mu$m detection. This 6\,GHz emission has no $870\,\mu$m continuum counterpart, but has a resolved optical/near-IR counterpart whose colour is similar to that of the SMG, as well as a confirmed \co\ \jthreetwo\ detection at the same redshift as the SMG \citep{wardlow18}. While radio jets have been found to be capable both of driving outflows of molecular gas from an AGN host \citep[e.g.\ ][]{dasyra15}, and of triggering star formation where they collide with the ISM of nearby galaxies \citep[e.g.\ ][]{lacy17}, the co-existence of strong radio and molecular gas emission, along with a bright optical counterpart that is clearly distinct from the SMG suggests that this radio feature is unlikely to be a jet originating in AS2UDS\,0017.1, but is instead a separate star-forming companion galaxy at the same redshift. A detailed analysis of this source is presented in \citet{wardlow18}, however for the present work, we note the complex radio morphology, and fit the 6\,GHz thumbnail with two Gaussian components. The source size quoted in Table\,\ref{tab:properties} is that of the component aligned with the $870\,\mu$m emission only.

\end{document}